\documentclass[nofootinbib,prd,aps]{revtex4}

\usepackage{amssymb}
\usepackage{amsfonts}
\usepackage{graphicx}

\def \be{\begin{equation}}
\def \ee{\end{equation}}
\newcommand{\field}[1]{\mathbb{#1}}
\def\boxit #1{\vbox{\hrule\hbox{\vrule\kern 2 pt \vbox {\kern 2 pt #1 \kern 2 pt} \kern 2 pt \vrule }\hrule}}

\begin{document}
%\topmargin = 0.3 true in   % seems to be required....   conflicts with arxiv...

\title{A Primer on Functional Methods and the Schwinger-Dyson Equations}

%\classification{11.15.Tk, 11.10.Ef}
%\keywords      {Schwinger-Dyson Equations, Functional Methods}

\author{Eric S. Swanson}

\affiliation{Department of Physics and Astronomy, University of Pittsburgh, Pittsburgh PA 15260.}

\begin{abstract}
An elementary introduction to functional methods and the Schwinger-Dyson equations is presented. Emphasis is placed on practical topics not normally covered in textbooks, such as a diagrammatic method for generating equations at high order, different forms of Schwinger-Dyson equations, renormalisation, and methods for solving Schwinger-Dyson equations.
\end{abstract}

\maketitle

\tableofcontents

\section{Introduction}

These notes introduce the derivation and uses of Schwinger-Dyson equations (and nonperturbative methods in general) to students who have some  familiarity with quantum field theory. These methods are based on the functional approach to quantum field theory and the path integral, and draw heavily on analogies to statistical mechanics. Thus exposure to these concepts will be beneficial to the student.

The path integral for quantum mechanics is reviewed very briefly and then applied to fields. This leads to a short introduction to functional calculus, diagrammatics, and the effective action. The power of the functional method is illustrated with a simple demonstration of Goldstone's theorem for the O(2) sigma model. An aside on the functional Hamiltonian and the variational principle follows. We then develop the Schwinger-Dyson master equation in a convenient form and derive a diagrammatic method to generate Schwinger-Dyson equations. Applications to  $\varphi^4$ theory, a contact fermion model, and QED follow. Short discussions of renormalisation, approximation schemes, and numerical techniques are included.

Although these notes are introductory, the discussion will occasionally get technical. Skipping such technical sections will do no harm to understanding the material. 
If you are familiar with things like generating functionals and 1PI diagrams then you can just skip to section \ref{schwingerdyson}.

\section{Nonperturbative Quantum Field Theory}

One often hears the word `nonperturbative' in relation to quantum field theory, and especially Quantum Chromodynamics. But its meaning is sometimes distorted and it is worthwhile to dwell on it for a moment.

In the context of quantum field theory, a nonperturbative property is one that {\it cannot} be obtained in perturbation theory. Notice that this does {\it not} mean that a 
 coupling constant is large. For example,
the series

\begin{equation}
F = 1 + x + x^2 + \ldots
\label{F}
\end{equation}
does not converge for $|x| > 1$, however one can formally sum it to obtain

\be
F = \frac{1}{1-x}
\ee
and then use this as a {\it definition} of $F$ for all $x$.

But consider the function

\be
F = {\rm e}^{-1/x}.
\ee
The Taylor series for this function about the origin is

\be
F = 0 + 0 + 0 + \ldots,
\ee
which is not terribly informative. Thus if the only tool one had to examine `F-theory' were perturbation theory about $x=0$, no progress could be made. We say that F-theory is nonperturbative. 

There is a common impression that being nonperturbative is akin to a death knell for a theory, consigning it to a class of theories that can only be examined with numerical methods. This is overly pessimistic, in the case of F-theory a 
sufficiently acute student might examine it in a new way and determine that it obeys the condition

\be
x^2\, \frac{dF}{dx} = F.
\ee
This is an easy equation to solve and tells us everything about $F$, up to a constant.

Perturbation theory is often the only tool a student learns to compute with quantum field theory. But as we will see, many field theories are nonperturbative and other methods must be employed. As with F-theory, a declaration that a certain problem is nonperturbative should not be considered fatal. Rather it is an indication that creativity is required.

%
%Examples of nonperturbative phenomena in quantum field theory are
%
%$\bullet\ $ dynamical symmetry breaking, such as chiral symmetry, magnetisation, and the Higgs phenomenon
%
%$\bullet\ $ quark and gluon confinement, the Wilson loop potential of QCD
%
%$\bullet\ $ mass generation in gauge theories
%

\section{Scalar Field Theory}

Following long tradition, we shall focus on the properties of scalar field theory in most of these notes. More realistic theories will be considered in section \ref{some}.

Scalar field theory is a relativistic model of self-interacting spin-zero particles. The `self-interacting' might be a bit odd for students accustomed to particles interacting via vector boson exchange; in this case forces are generated by scalar boson exchange, and there are no `matter' particles. But interesting things can still happen: `phions' could bind into multiphion particles, parity symmetry could be broken, particle interactions could be screened to the extent that they disappear at moderate distances.

A free phion is described with the Lagrangian 

\be
L_0 = \int d^{(d-1)}x\, \left( \frac{1}{2}\partial_\mu\varphi \partial^\mu \varphi - \frac{1}{2}m_0^2 \varphi^2\right).
\ee
Self interactions are introduced via an interaction term $L = L_0-\int V$. This is often taken to be 

\be
V = \frac{\lambda_0}{24} \varphi^4.
\ee
One also sees interactions of the type $V = g \varphi^3$ but this theory does not have a finite classical vacuum (as can be seen by letting $\varphi$ be a constant and considering $\varphi <0$), so we do not consider it here.
Thus our classical action is

\begin{equation}
S = \int d^dx \, \left( \frac{1}{2}\partial_\mu \varphi \partial^\mu \varphi - \frac{1}{2} m_0^2 \varphi^2 - \frac{\lambda_0}{24} \varphi^4 \right).
\end{equation}

Some remarks on the notation: 

(i) We work in $d$ spacetime dimensions. Since (with $\hbar = 1$) the action is unitless, one obtains

\begin{eqnarray}
[m_0] &=& 1 \nonumber \\
{[}\varphi{]} &=& -1+d/2 \nonumber \\
{[}\lambda_0{]} &=& 4-d
\label{unitsEq}
\end{eqnarray}
where units are measured in terms of energy. Thus $[m_0] = 1$ means that this parameter can be specified in GeV.

(ii) At this stage, the meaning of the parameters $m_0$ and $\lambda_0$ is not known. One must compute observable properties, such as the mass of a phion molecule or a phion-phion scattering cross section, to enable an interpretation of the parameters of the theory. A noninteracting phion obeys the dispersion relation $E^2 = p^2 + m_0^2$ so it is tempting to interpret $m_0$ as the phion mass. But it must be remembered that this is only an approximation, interactions will change this relationship, and hence the interpretation of $m_0$. For this reason, $m_0$ and $\lambda_0$ are called the {\it bare parameters} of the theory. The physical phion mass, $m$, is related to $m_0$ and $\lambda_0$ in a complicated way. (A similar statement holds for the coupling $\lambda$).

A famous property of quantum field theory is that calculations generate infinite expressions. The process of {\it renormalisation} absorbs these infinities in the bare parameters, leaving well-defined finite physical parameters, $m$ and $\lambda$. Renormalisation in the context of Schwinger-Dyson equations and the functional method will be discussed in Section \ref{renormalisation}. For now we note
that renormalisation would be required even for a theory with no infinities because, as we have stressed, $m$ and $m_0$ are not related in a simple way.

The classical field equations for $\varphi^4$ theory are

\be
(\partial^2 + m_0^2)\varphi + \frac{\lambda_0}{6} \varphi^3 = 0.
\ee
This is a nonlinear relativistic field equation that must already be solved with numerical methods or inspired guesswork. Equations of this sort can often yield surprising phenomena such as dispersionless waves (solitons).  Rich nonperturbative phenomena in the quantum realm should be expected!

\section{Functional Methods}

\subsection{Introduction}

Following Dirac, canonical quantisation of $\varphi^4$ theory proceeds by replacing oscillator strengths in the Fourier transform of $\varphi$ with ladder operators. We summarise this procedure with the expression

\be
\varphi \to \hat \varphi.
\ee
Scattering amplitudes can be computed with the aid the Gell-Mann-Low equation:

\be
\langle {\rm out}| S | {\rm in}\rangle = \frac{\langle {\rm out} | T\, {\rm e}^{-i\int_{-T}^{T} V_I(t) dt}|{\rm in}\rangle}{\langle 0 | T\, {\rm e}^{-i \int V_I}|0\rangle}.
\label{gmlEq}
\ee
Evaluating the right hand side requires computing matrix elements of products of fields, for example $\langle {\rm out} \vert \int d^dx\, \hat \varphi^4(x) | {\rm in} \rangle$. This laborious exercise in ladder algebra is greatly simplified with the aid of Wick's theorem:

\be
T[\hat \varphi(x)\hat \varphi(y)] = N[\hat \varphi(x)\hat \varphi(y)] + \langle 0 | \hat \varphi(x) \hat \varphi(y) | 0 \rangle.
\label{wickEq}
\ee
Besides saving labour, Wick's theorem permits computations solely in terms of field operators, rather than Fourier components of fields. Furthermore, practice with Eq. \ref{wickEq} reveals that the algebra of Wick's theorem is the same as that of derivatives. These seemingly trivial observations carry great weight because they hint that it is possible to quantise field theory solely in terms of the functional properties of fields. This forms the basis of {\it functional quantisation}.

\subsection{Path Integral -- Quantum Mechanics}

Functional quantisation takes the Feynman path integral as its starting point. I assume that the reader has been exposed to the path integral before, but will nevertheless take this opportunity to review some, possibly lesser known, aspects of this method.

 The path integral is interpreted as a sum of phases associated with all possible paths that a system can take from initial to final state. In particular, a 
transition amplitude is written as

\be
 \langle x_f;t_f \vert x_i;t_i \rangle = \int {\cal D}x \, 
e^{{i \over \hbar} \int_{t_i}^{t_f}dt {\cal L}(\dot x,x)}.
\label{formalFPIEq}
\ee
However, as it stands this expression is meaningless. In fact a better way to write 
it is

\be
 \langle x_f;t_f \vert x_i;t_i \rangle = \sum e^{{i \over \hbar}S_{cl}(path)} 
\ee
where the sum extends over all paths connecting $(x_i,t_i)$ with $(x_f
,t_f)$. This is traditionally represented by discretising the time 
axis into slices of size $\epsilon$ and integrating over all $x_n = x(n \epsilon)$. 
Notice however that this means that we are no longer summing over 
all paths but only those which obey the order $1/2$ Lipschitz 
condition:

\be
  \vert x(t') - x(t) \vert < k \vert t' - t \vert^{1/2}
\label{lipsEq}
\ee
where $k$ is some constant. 

Thus we rewrite Eq. \ref{formalFPIEq} in a more precise form as

\be
\langle x_f;t_f \vert x_i;t_i \rangle = \lim_{\epsilon \to 0} \int 
\prod_{j=1}^N {dx_j \over A} \, e^{{i \over \hbar} 
 \sum_{j=0}^N S_{cl}(x_j \rightarrow x_{j+1})} 
\ee
where $S_{cl}(x_j \rightarrow x_{j+1})$ is the classical action 
evaluated between $x(j\epsilon)$ and $x((j+1) \epsilon)$, $t_0 = t_i$, 
and $ t_{N+1} = t_f$. 

For a free particle the classical path is given by
 
\be
 x_{free}(t) = { x_{j+1} - x_j \over \epsilon} (t - t_j) + x_j 
\label{classicalPathEq}
\ee
so that 

\be
\langle x_f;t_f \vert x_i;t_i \rangle = \lim_{\epsilon \to 0} \int 
\prod_{j=1}^N {dx_j \over A} \, e^{{i m \over 2 \hbar
\epsilon} \sum_{j=0}^N (x_{j+1} - x_j)^2 } .
\ee
If the particle is moving in a potential we do not know the classical 
path in general, however this poses no problem as we shall see.

For a particle in a harmonic potential

\be
 {\cal L} = {1 \over 2} m \dot x^2 - {1 \over 2} m \omega^2 x^2 
\ee

\noindent
the classical path is

\begin{eqnarray}
x_{sho} &=& x_i \cos \omega (t-t_i) + { x_f - x_i \cos \omega 
\epsilon \over \sin \omega \epsilon }  \sin \omega (t-t_i) \nonumber \\
\phantom{x} &\simeq& x_{free} + O[(t-t_i)^2] \nonumber \\
\phantom{x} &=& x_{free} + O(\epsilon^2)
\end{eqnarray}
We see that the classical path differs little from the free path after 
short durations. In general $x_{cl} = x_{free} + O(\epsilon)$ and

\be
 \int_{t_i}^{t_f} V(x_{cl}) dt =  \int_{t_i}^{t_f} V(x_{free} + O(
\epsilon)) dt = \epsilon V(x_i) + O(\epsilon^{3/2})
\ee

\noindent
upon use of Eq. \ref{classicalPathEq}. The order $\epsilon^{3/2}$ term does not 
contribute to time evolution of the system so that it is valid to 
take $x_{cl} = x_{free}$ between the points $x_j$ and $x_{j+1}$ and to 
evaluate $V$ at some point in the interval (since the difference is higher 
order in $\epsilon$). The end result is that it is valid to discretise as 
follows

\be
S_{cl}(x_j \rightarrow x_{j+1}) = \int_{t_i}^{t_f} \left[ {1 \over 2} m 
\dot x^2 - V(x) \right] \, dt = {m \over 2 \epsilon} (x_{j+1} - 
x_j)^2 - \epsilon V({x_j + x_{j+1} \over 2}) + O(\epsilon^{3/2}) 
\label{SdefEq}
\ee

The temporal evolution of a wavefunction is given by

\be
 \psi(x;t+\epsilon) = \int {dy \over A} e^{{i \over \hbar}S_{cl}(y 
\rightarrow x)} \, \psi(y;t).
\label{psidotEq}
\ee
Thus for a free particle

\be
\psi(x;t+\epsilon) = \int {dy \over A} e^{{i m \over 2 \hbar 
\epsilon} (x-y)^2} \, \psi(y;t) .
\ee
The integral is washed out for large values of $x-y$ since 
the exponential oscillates rapidly in that region. In fact the only 
appreciable contribution is when $x-y \sim \sqrt{\epsilon}$. Now let 
$ \eta = x-y$ and expand in $\epsilon = \eta^2, \eta^4/\epsilon, 
\ldots$ to get

\begin{eqnarray}
\psi(x;t) &=& \int_{-\infty}^\infty \, {d \eta \over A}
\, \psi(x;t) \, e^{{i m \over 2 \hbar \epsilon} \eta^2}  , \quad \rightarrow
 A = \sqrt{{ 2 \pi i \epsilon \hbar \over m}}  \nonumber \\
\epsilon \dot \psi &=& { i \hbar \epsilon \over 2 m} \psi'', \quad \rightarrow i \hbar
\dot \psi = {- \hbar^2 \over 2m} \psi'' \nonumber \\
\epsilon^2 \ddot \psi &=& {-\hbar^2 \epsilon^2 \over 4 m^2} \psi^{IV}, 
\quad {\rm etc.} 
\end{eqnarray}

\noindent
Thus the zero order term fixes the normalization, the first order 
term yields the Schr\"odinger equation, and the higher order terms are 
simply multiple applications of the Schr\"odinger equation, and so 
represent no new information. This technique can be used to obtain the quantum mechanical Hamiltonian associated with a given system.

\subsection{Path Integral -- Quantum Field Theory. Functional Calculus}

The generalisation to field theory proceeds by discretising spacetime, $\varphi(x) \to \varphi_i$, and defining the sum over paths as

\be
\int D\varphi \to \int \prod d \varphi_i.
\ee

Continuing with the development will very quickly lead to quantities like $\delta(\varphi_i - \varphi_j)$, $\partial/\partial \varphi_i$, and functions $f(\{\varphi_i\}) \mapsto \field{C}$. The continuum version of a function like $f$ maps elements of a function space to complex space; these are called a {\it functionals} and are denoted $f[\varphi]$. A simple example would be 

\be
f[\varphi] = \int ds\, \sqrt{1+(\frac{d}{ds}\varphi(s))^2}
\ee
(this is the arc length functional). Manipulating functionals is called {\it functional analysis} or 
{\it functional calculus}. Unfortunately, functional methods are not nearly as well developed as real analysis or calculus, even defining functional integrals is fraught with mathematical complexities.But we are physicists and will happily ignore anything that gets in the way of progress. The {\it functional derivative} is introduced in analogy to the integral:

\be
\frac{\partial}{\partial \varphi_i} \varphi_j = \delta_{ij} \to \frac{\delta}{\delta \varphi(x)} \varphi(y) = \delta(x-y).
\label{f-derivEq}
\ee

Now consider the problem of minimising a functional $F[\varphi]$\footnote{Of course this is close to a physicist's heart: all classical physics follows from the minimisation of an action functional},

\be
F[\varphi] = \int f(\varphi,\partial_\mu \varphi).
\ee
Set $\varphi = \varphi+\delta \varphi$, substitute and extract the leading change in $F$:
\begin{eqnarray}
\delta F &=& \int \left[f(\varphi+\delta\varphi,\partial_\mu (\varphi+\delta\varphi)) - f(\varphi,\partial_\mu\varphi) \right] \nonumber \\
&=& \int \left[ \frac{\partial f}{\partial \varphi} \delta\varphi + \frac{\partial f}{\partial (\partial_\mu \varphi)} \partial_\mu \delta \varphi\right] \nonumber \\
&=& \int \left[ \frac{\partial f}{\partial \varphi} - \partial_\mu \frac{\partial f}{\partial (\partial_\mu \varphi)}\right] \delta \varphi.
\end{eqnarray}
The variation is arbitrary so the integrand must be zero at the minimum and we derive the Euler-Lagrange equations of motion (if $f$ is the Lagrangian density and $F$ is the action):

\be
\frac{\partial f}{\partial \varphi} = \partial_\mu\frac{\partial f}{\partial (\partial_\mu \varphi)}.
\ee
Now consider the same problem from the functional point of view: upon discretising the minimum (extremum) of $F[\varphi]$ is determined by the set of equations 

\be
\frac{\partial}{\partial \varphi_i} F(\{\varphi_j\}) = 0.
\ee
In the continuum this is 
\be
\frac{\delta}{\delta \varphi(x)}F[\varphi] = \frac{\delta}{\delta\varphi(x)}\int dy\, f(\varphi(y),\partial_\mu \varphi(y)) = 0
\ee
Taking the derivative, integrating by parts, and using Eq. \ref{f-derivEq} gives the Euler-Lagrange equations of motion. This is nice -- the functional derivative provides a  compact and transparent way to obtain the Euler-Lagrange equations of motion.

For practice consider the equations of motion of $\varphi^4$ theory:

\begin{eqnarray}
\frac{\delta}{\delta \varphi(y)} S[\varphi] &=& 
\frac{\delta}{\delta \varphi(y)} \int d^dx\, \left[ \frac{1}{2} \partial_\mu\varphi \partial^\mu \varphi - \frac{1}{2} m_0^2 \varphi^2(x) - \frac{\lambda_0}{24}\varphi^4(x)\right]  \nonumber \\
&=& \int d^d x\, \left[ 2 \frac{1}{2} \frac{\delta \varphi(x)}{\delta \varphi(y)}(-\partial^2 \varphi(x)) - m_0^2 \varphi(x) \delta(x-y) - \frac{\lambda_0}{24}\cdot 4 \cdot \varphi^3(x) \delta(x-y) \right] \nonumber \\
&=& -(\partial_y^2 + m_0^2)\varphi(y) - \frac{\lambda_0}{6}\varphi^3(y) 
\label{eomEq}
\end{eqnarray}

We now return to the path integral over fields.
One can write the amplitude for a state $\varphi_i$ at $-T$ to evolve to $\varphi_f$ at $T$ as

\be
\langle \varphi_f(x;T) | {\rm e}^{-iH\cdot 2T} | \varphi_i(x;-T)\rangle = \int D\varphi|_{\varphi(x,-T)=\varphi_i(\vec x)}^{\varphi(x,T)=\varphi_f(\vec x)} \, {\rm e}^{i\int_{-T}^{T} {\cal L}}
\ee
If one considers the limit $T\to \infty$ and assumes {\it adiabatic switching}\footnote{This means that the initial and final states evolve to ground states of the noninteracting system.} then the expression becomes manifestly covariant and represents
the vacuum-to-vacuum transition amplitude. In the presence of an external source field, $J$, this is called the {\it generating functional}. For a scalar field theory

\begin{equation}
Z[J] = \int D\varphi\, {\rm e}^{i S + i\int d^4x\, J(x) \varphi(x)}
\label{gfcnlEq}
\end{equation}

\noindent
$Z$ is called the generating functional because it contains all of the information necessary to evaluate observables in scalar field theory.
For example, it can be shown that the two point function defined by

\be
\langle \varphi_1 \varphi_2\rangle = \frac{\langle 0 | T[ \varphi_{I}(x_1) \varphi_I(x_2) {\rm e}^{-i \int V_I}] | 0\rangle}{\langle 0 | T[{\rm e}^{-i \int V_I}]|0\rangle} 
\ee
is also given by
\be
\langle \varphi_1 \varphi_2\rangle = 
\frac{1}{Z} \int D\varphi\, \varphi_1\varphi_2\, {\rm e}^{iS + i \int J\varphi}|_{J=0}.
\ee

We check this explicitly at zeroth order in perturbation theory:

\be
\langle \varphi_1\varphi_2\rangle = i \int \frac{d^4 k}{(2\pi)^4} \frac{{\rm e}^{-i k\cdot(x_1-x_2)}}{k^2-m^2+i\epsilon} \equiv \Delta(x_1-x_2).
\label{deltaEq}
\ee

Comparing to the functional method requires knowledge of the 
quadratic integral

\be
\int D\varphi\, {\rm e}^{\frac{i}{2} \int\int \varphi_1 M_{12} \varphi_2 + i \int J\varphi} = ({\rm det} \, M)^{-1/2} {\rm e}^{-\frac{i}{2}\int\int J_1 (M^{-1})_{12}J_2}.
\ee

Use the last equation to confirm the expression for the free propagator:

\begin{eqnarray}
\langle \varphi_1\varphi_2\rangle &=& \frac{1}{Z_0} \frac{\delta}{i\delta J_1}\frac{\delta}{i \delta J_2} Z_0[J]|_{J=0} \nonumber \\
 &=& -i (\frac{1}{\partial^2 + m^2})_{12} \nonumber \\
&=& i \int \frac{d^4 k}{(2\pi)^4} \frac{{\rm e}^{-i k\cdot(x_1-x_2)}}{k^2-m^2+i\epsilon}.
\label{pertPropEq}
\end{eqnarray}
We have defined the free generating functional $Z_0$, which is obtained from Eq. \ref{gfcnlEq} upon setting the interaction to zero.
Notice that the Feynman prescription was introduced in the last step of Eq. \ref{pertPropEq}. It is of course required to select the appropriate poles in the Feynman propagator. This can be accomplished by simply setting  

$$
\partial^2 + m^2 \to \partial^2 + m^2  - i \epsilon
$$
in the action. This has the effect of multiplying the action by ${\rm e}^{-\epsilon VT}$, which acts as a convergence factor on what is otherwise a distressingly oscillatory integrand.

For later we note that the propagator is a Greens function of the Klein-Gordon equation:

\begin{equation}
(\partial_x^2 + m^2) \Delta(x-y) = -i \delta^4(x-y).
\label{GfcnEq}
\end{equation}

%\subsection{Connected and One-particle-irreducible Diagrams}
\subsection{Diagrammatics}
\label{diagrammatics}

\begin{figure}[ht]
\includegraphics[width=7cm]{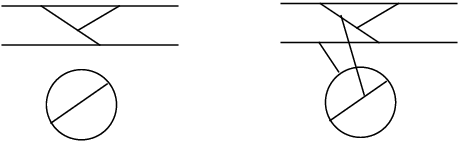}
\caption{(left) a disconnected diagram, (right) a connected diagram.} 
\label{connFig}
\end{figure}

Recall that the Gell-Mann-Low equation (Eq. \ref{gmlEq}) is normalized with a vacuum matrix element. This has the effect of eliminating all bubble diagrams from scattering matrix elements. This is desirable since bubble diagrams do not contribute to scattering and they introduce factors of the spacetime volume (ie, infinity) into the formalism.  Diagrams that contain portions that do not connect to scattering particles (external legs) are called {\it disconnected} (see Fig. \ref{connFig}).

In our case we must divide all expressions by $Z$ (as we have been doing above),
or equivalently, consider derivatives of $\log Z$. This is most easily achieved by defining the {\it connected generating functional}:

\begin{equation}
{\rm e}^{iF[J]} = Z[J].
\label{FdefEq}
\end{equation}
Derivatives of $F$ automatically yield connected Greens functions.

\begin{figure}[ht]
\includegraphics[width=7cm]{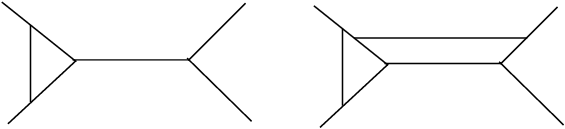}
\caption{(left) not 1PI, (right) 1PI}
\label{1piFig}
\end{figure}

A {\it one-particle-irreducible} (1PI) diagram (also sometimes called a {\it proper} diagram)
is a connected diagram that remains connected when any internal line is removed (Fig. \ref{1piFig}).

\noindent
1PI diagrams are useful because they can be summed in simple ways to generate full diagrams. For example if an open circle represents a 1PI vertex and a full circle represents the analogous full diagram then
the two point function is related to the two point 1PI diagram by a geometric sum (see Fig. \ref{SDE2}):

\begin{figure}[ht]
\includegraphics[width=8cm]{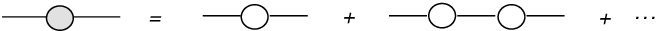}
\caption{Full two-point function in terms of the two-point 1PI diagram.}
\label{SDE2}
\end{figure}

\begin{figure}[ht]
\includegraphics[width=7cm]{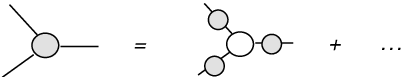}
\caption{Full three-point function in terms of the three-point 1PI diagram.}
\label{SDE2b}
\end{figure}

Similarly, the full three point function is given in terms of 1PI three point functions (see Fig. \ref{SDE2b}).

\subsection{The Legendre Transformation}
\label{legendre}

Unfortunately, obtaining 1PI diagrams from the generating functional is not as simple as performing a normalisation. However the analogy to statistical mechanics helps again, and it is possible to construct a 1PI generating functional in analogy to the Gibbs free energy. To illustrate, consider a spin system coupled to an external magnetic field, $\vec B$. Call the induced magnetisation $\vec M$. The Gibbs free energy is defined in terms of the free energy by

\be
G = F - MB
\ee
The statistical mechanical generating function is  
\be
Z[B] = {\rm e}^{-\beta F} = {\rm tr} \, {\rm e}^{-\beta H(S) - \beta B\cdot S}
\ee
from which one obtains
\be
M = \langle S\rangle \nonumber = \frac{{\rm tr} \, S \, {\rm e}^{-\beta H}}{{\rm tr}\, {\rm e}^{-\beta H}} = -\frac{1}{\beta} \frac{\partial}{\partial B} \log Z\vert_{B=0} = \frac{\partial F}{\partial B}
\label{GibbEq}
\ee

Now compute the change in the Gibbs free energy with respect to the induced magnetisation:
\begin{eqnarray}
\frac{\partial G}{\partial M} &=& \frac{\partial F}{\partial M} - B - M \frac{\partial B}{\partial M} \nonumber \\
&=& \frac{\partial F}{\partial B}\frac{\partial B}{\partial M} - B - M \frac{\partial B}{\partial M} \nonumber \\
&=& -B
\end{eqnarray}
where we have used Eq. \ref{GibbEq}. The preferred thermodynamic configuration is determined by the minimum of $G(M)$ and all quantities can be derived in terms of the induced, rather than the applied, magnetic field.

The procedure is called the {\it Legendre transformation} in quantum field theory.
In analogy, we shall define the {\it  classical field} as the vacuum expectation value of the field operator in the presence of a source:

\begin{equation}
\varphi_{cl}(x) \equiv \frac {\langle \varphi(x)\rangle}{\langle 1 \rangle} = \frac{1}{Z[0]} \frac{\delta}{i\delta J(x)} Z[J] = \frac{\delta}{\delta J(x)} F[J]
\label{phicl-defn}
\end{equation}
Notice that $\varphi_{cl}$ is an implicit function of the source.
Computing with the free action we see

\begin{equation}
\varphi_{cl}^{(0)}(x)= i \int d^dy\,\Delta(x-y) J(y).
\end{equation}
%We have introduced a notation that will be employed often in the subsequent discussion: namely continuous variables such as $x$ and $k$ as being discrete (imagine placing the theory onto a discrete spacetime grid) and Einstein summation convention is employed:
%
%$$
%a_x b_x = \int d^d x \, a(x) b(x)
%$$
%
%$$
%a_k b_k = \int \frac{d^dk}{(2\pi)^d} a(k) b(k).
%$$
%We shall flip between notations freely in the following.

Using Eq. \ref{GfcnEq} we obtain

$$
(\partial_x^2 + m^2) \varphi_{cl}^{(0)}(x) = J(x)
$$
This equation tells us that it is possible to replace references of the source with references to the classical field (by appropriately  inverting $\partial^2 + m^2$). It is convenient to make this replacement implicitly via the Legendre transformation:

\begin{equation}
\Gamma^{(0)}[\varphi_{cl}^{(0)}] = F^{(0)}[J] - \int d^dx\,J(x) \varphi_{cl}^{(0)}(x)
\end{equation}

\boxit{\begin{quote}
\noindent
verify that $\Gamma^{(0)}$ is independent of $J$. 
\vskip .3 cm
\noindent
Ans: take $\frac{\delta}{\delta J}$ and use the definition of $\varphi_{cl}$.
\end{quote}}

\boxit{\begin{quote}
\noindent
evaluate $\Gamma^{(0)}$. 
\vskip .3 cm
\noindent
Ans: substitute , integrate by parts, and use Eq. \ref{GfcnEq}.
\end{quote}}

You should find that the zeroth order expression for $\Gamma$ is

\begin{equation}
\Gamma^{(0)}[\varphi_{cl}] = \frac{1}{2} \int d^d x \, \left( \partial_\mu \varphi_{cl}\partial^\mu \varphi_{cl} - m_0^2 \varphi_{cl}^2 \right)
\end{equation}
Thus $\Gamma^{(0)}$ is the free classical action, and we call $\Gamma$ the {\it  effective action}. 

All of this is generalised to the interacting case in a natural way:

\begin{eqnarray}
\varphi_{cl} &\equiv& \frac{\delta}{\delta J} F[J] 
\label{FEq} \\
\Gamma[\varphi_{cl}] &\equiv& F[J] - \int d^d x\, J(x) \varphi_{cl}(x) 
\label{G-eqn} \\
\frac{\delta \Gamma}{\delta \varphi_{cl}(x)} &=& - J(x) 
\label{J-eqn}
\end{eqnarray}
The last equation is obtained by taking the derivative of the second and recalling that $J$ and $\varphi_{cl}$ are independent variables.
As alluded to above, the effective action is a generating functional for 1PI diagrams.
Specifically, the expansion of the effective action is in terms of 1PI n-point functions:

\be
\Gamma[\varphi_{cl}] = \sum_{n=2}^\infty \frac{1}{n!} \int d^dx_1 \ldots d^dx_n\,\Gamma^{(n)}(x_1,\ldots,x_n) \, \varphi_{cl}(x_1) \cdots \varphi_{cl}(x_n).
\label{G-expansionEq}
\ee

\subsection{Evaluating and Interpreting the Effective Action}
\label{evaluating}

\subsubsection{The Effective Potential}

If the Hamiltonian is $T+V$ and the action is $T-V$, is there an analogue to the interaction, $V$ in the {\it effective} action? The answer is yes, it is called the {\it effective potential} and it can be obtained by 
performing a derivative expansion of the effective action:

\be
\Gamma(\varphi_{cl}) = \int d^dx\, \left( - V_{eff}(\varphi_{cl}) + \frac{1}{2} Z_{eff}(\varphi_{cl}) \partial_\mu \varphi_{cl} \partial^\mu \varphi_{cl}  + \ldots\right).
\label{derExpansionEq}
\ee
Notice that $V_{eff}$ and $Z_{eff}$ are functions of $\varphi_{cl}$, not functionals. Thus

\be 
V_{eff}(\varphi) = V_0 + V_2 \varphi^2 + V_4 \varphi^4 + \ldots
\label{VExpansionEq}
\ee
with a similar equation for $Z_{eff}$. This expression implies that it is sufficient to take $\varphi_{cl}$ equal to a constant to obtain the effective potential. This simplifies the 1PI expansion of the effective action (Eq. \ref{G-expansionEq})

\be
\Gamma[\varphi_{cl}=const] = \sum_{n=2}^\infty \frac{1}{n!} \int d^dx_1 \ldots d^dx_n\,\Gamma^{(n)}(x_1,\ldots,x_n) \, \varphi_{cl}^n
\ee
or
\be
\Gamma(\varphi_{cl}=const)= \sum_{n=2}^\infty \frac{1}{n!} \tilde\Gamma^{(n)}(0,\ldots,0) \, \varphi_{cl}^n
\ee
where momentum space 1PI vertex functions have been introduced

\be
\Gamma(x_1,\ldots,x_n) = \int \frac{d^dp_1}{(2\pi)^d} \ldots \frac{d^dp_n}{(2\pi)^d} {\rm e}^{-i p_1\cdot x_1 \ldots -i p_n \cdot x_n} \, (2\pi)^4\, \delta(p_1+\ldots+ p_n) \tilde\Gamma(p_1,\ldots,p_n).
\ee
We conclude that {\it the effective potential is determined by Feynman graphs evaluated at zero external momentum}.

We obtain an interpretation of the effective potential by recalling that the generating functional is also a vacuum-to-vacuum transition element. The external source distorts all eigenstates of the Hamiltonian so we shall label states $\vert n(J)\rangle$ in the following. Note that if the source is time-dependent then the incoming vacuum need not be the same as the outgoing. We eliminate this possibility by considering $J=J(\vec x)$. Thus

\begin{eqnarray}
Z[J] &=& \frac{1}{Z[0]} \langle \varphi_f(\vec x)| {\rm e}^{-iH(J)T}|\varphi_i(\vec x)\rangle \nonumber \\
&=& \frac{1}{Z[0]} \sum_n{\rm e}^{-iE_n(J) T} \langle\varphi_f|n(J)\rangle\, \langle n(J)|\varphi_i\rangle \nonumber \\
&\to& {\rm e}^{-i(E_0(J)-E_0)T} \frac{\langle \varphi_f|0(J)\rangle\, \langle 0(J)|\varphi_i\rangle}{\langle \varphi_f|0\rangle\,\langle 0|\varphi_i\rangle} \nonumber \\
&\equiv& {\rm e}^{iF}.
\end{eqnarray}
A complete set of states has been inserted in the second step and the large time limit has been taken in the third step.
Thus (take $E_0=0$ and neglect the normalisation term)
\be
F[J] = -\langle 0(J)|\left[H - \int J\varphi_{cl}\right] |0(J)\rangle \cdot T.
\ee
Recalling the definition of the effective action (Eq. \ref{G-eqn}) then gives

\be
\Gamma[\varphi_{cl}] = -\langle 0(J)|H|0(J)\rangle \cdot T.
\ee

Now let the classical field be a constant and refer to the derivative expansion of Eq. \ref{derExpansionEq} to obtain

\be
V_{eff}(\varphi_{cl}=const) = \langle 0; \varphi_{cl}|{\cal H}|0; \varphi_{cl}\rangle
\ee
where the notation for the states indicates that the vacuum configuration depends on the source, and hence on $\varphi_{cl}$.   The spacetime integral in equation $\ref{derExpansionEq}$ yields a factor of $VT$; dividing by the spatial volume then gives the Hamiltonian density that appears on the right hand side.

Thus there is a simple and natural relationship  between the effective potential and the Hamiltonian of the system. From this we can derive that the effective potential must be real since the Hamiltonian is Hermitian\footnote{Perhaps things are not so simple. Recall that scattering states are non-normalisable and hence make the Hamiltonian non-hermitian. Could a similar thing happen in field theory? One could, for example, interpret such an occurrence as an indication of a  {\it metastable vacuum}.}. 
Furthermore, in analogy with statistical mechanics, it can be shown that the effective potential must be convex.
Lastly, recall that $\delta \Gamma/ \delta \varphi_{cl} = -J$ so that the value of the (constant) classical field is determined by the simple equation

\be
\frac{\partial V_{eff}}{\partial \varphi_{cl}}|_{J=0} = 0.
\label{V-derivEq}
\ee

\subsubsection{The Loop Expansion}

[This section is a simplified version of chapter 11 of Ref. \cite{PS}]

At tree order the effective potential is given by $V_{eff} = \frac{1}{2} m_0^2 \varphi_{cl}^2 + V(\varphi_{cl})$. 
Quantum corrections can be computed in a clear way with the functional formalism that is called the {\it loop expansion}.

For example, consider the computation of $Z[J]$  for $\varphi^4$ theory to one-loop
order. Let 

\be
\varphi = \varphi_{cl} + \eta
\ee
where the classical vacuum satisfies

\be
\frac{\delta S}{\delta \varphi}|_{\varphi=\varphi_{cl}} = 0.
\ee
Expand the action

\be
\int {\cal L}(\varphi_{cl} + \eta) = \int d^d x {\cal L}(\varphi_{cl}) + \frac{1}{2} \int \eta(x) \frac{\delta^2 {\cal L}}{\delta \varphi(x) \delta\varphi(y)}\eta(y) +\ldots
\ee
Thus 

\be
Z[J] \approx {\rm e}^{i \int {\cal L}(\varphi_{cl}) + i \int J \varphi_{cl}} \cdot \int D\eta \, {\rm e}^{\frac{i}{2}\int \eta \frac{\delta^2 {\cal L}}{\delta\varphi\delta\varphi} \eta},
\ee
\be
Z[J] \approx {\rm e}^{i \int {\cal L}(\varphi_{cl}) + i \int J \varphi_{cl}} \cdot {\rm det}\left( \frac{\delta^2 {\cal L}}{\delta\varphi\delta\varphi}\right)^{-1/2}
\ee
The second factor represents quantum (one-loop) corrections to the classical effective potential. 

Continuing,

\be
iF \approx i \int d^d x {\cal L}(\varphi_{cl}) + i \int J \varphi_{cl} - \frac{1}{2} \log \det 
\left( \frac{\delta^2 {\cal L}}{\delta\varphi\delta\varphi}\right)
\label{F2Eq}
\ee
and $i\Gamma$ is the same expression without the source term. Now use the expression

\be
\log \,{\rm det}\, {\cal O} = {\rm tr}\,\log \, {\cal O}.
\ee

\boxit{\begin{quote}
Show this.
\vskip .3 cm
Ans. Diagonalise ${\cal O}$.
\end{quote}}

\noindent
to obtain
\be
\Gamma \approx \int {\cal L}(\varphi_{cl}) + \frac{i}{2} {\rm tr}\,\log 
\left( \frac{\delta^2 {\cal L}}{\delta\varphi\delta\varphi}\right).
\ee
Substitute
\be
\frac{\delta^2 {\cal L}}{\delta\varphi\delta\varphi} = -\partial^2 - m_0^2 - V''(\varphi_{cl})
\ee
and evaluate
\begin{eqnarray}
{\rm tr} \, \log 
(\partial^2 + m_0^2 + V''(\varphi_{cl})) &=& VT\cdot \int \frac{d^dk}{(2\pi)^d}\, \log(
-k^2 + m_0^2 + V''(\varphi_{cl})) \nonumber \\
&=& VT\cdot i \, \int \frac{d^dk_E}{(2\pi)^d} \, \log(k_E^2 + m_0^2 + V''(\varphi_{cl})) \nonumber \\
&=& i\frac{VT}{8\pi^2}\,\Big[ -\frac{1}{8}\Lambda^4 + \frac{1}{2}\Lambda^2\,(m_0^2+V'') + \nonumber \\
&& + \frac{1}{4} (m_0^2+V'')^2 \, \log\left(\frac{m_0^2+V''}{\Lambda^2+m_0^2+V''}\right)\Big]
\end{eqnarray}
In the second line we have made the {\it Wick rotation} $k_0 \to ik_d$. We have taken $d=4$ and regulated with a momentum cutoff to obtain the third line (for details on regulating field theory see section \ref{renormalisation})\footnote{The traditional method of computing has been presented. The observant student will note that the logarithm has units, which is unacceptable. Units can be supplied by subtracting $\log \Lambda^2$ while cutting off at $\Lambda$. This yields a term $\frac{1}{4}\Lambda^4\,\log(1+(m_0^2+V'')/\Lambda^2) \to \frac{1}{4}\Lambda^2(m_0^2+V'')$.}.

Remember that $\varphi_{cl}$ can be taken to be constant to obtain the effective potential. Substituting (Eq. \ref{derExpansionEq}) and cancelling the spacetime factors gives

\be
V_{eff}(\varphi_{cl}) = V(\varphi_{cl}) + \frac{1}{16\pi^2} \Big[ {\rm above}  \Big].
\ee

Unfortunately, the expression in brackets is ridiculously divergent. Dealing with divergences is called renormalisation. This topic will be discussed in section \ref{renormalisation}. For now we note that the $\Lambda^4$ term can be neglected because it contributes to the vacuum energy density, which we set to zero. Also recall that $m_0$ and $\lambda_0$ are bare parameters, and must be determined by comparing predictions to experiment. Doing so permits one to replace the bare parameters in terms of physical parameters, called $m$ and $\lambda$. These parameters are determined by the scale at which the experiment is made, which we call $M$. In practice one can set $V_{eff}''(\varphi_{cl}=0) = m^2$ and $V_{eff}''''(\varphi=M) = \lambda$ ($M$ is chosen as the renormalisation scale here because $M=0$ is awkward). For more information see Section \ref{renormV}. The end result is

\begin{eqnarray}
V_{eff}(\varphi_{cl}) &=& \frac{1}{2}m^2 \varphi_{cl}^2 + \frac{\lambda}{24}\varphi_{cl}^4 + \frac{1}{64\pi^2}\Big[ (V'')^2\, \log (\frac{V''}{m^2}) + \nonumber \\
&+& \frac{1}{2}\lambda m^2\varphi_{cl}^2 - \frac{25}{24} \lambda^2 \varphi_{cl}^4 + \frac{1}{4}\lambda^2 \varphi_{cl}^4\, \log\left(\frac{2m^2}{\lambda M^2}\right)\Big]
\label{VeffRenormEq}
\end{eqnarray}
where now $V''$ refers to the renormalised potential, $m^2\varphi^2/2 + \lambda\varphi^4/24$.

There are some subtleties to consider here

(i) the effective potential depends on the scale $M$. This scale arises when infinities are removed via the renormalisation procedure.
The $M$-dependence is absorbed into the definitions of the couplings, specifically if you choose a different scale, the couplings will change in such a way as to leave the potential invariant. This phenomenon goes by the name of the {\it renormalisation group}.

\boxit{\begin{quote}
Verify the renormalisation group.
\vskip .3 cm
Ans: Show that if a different scale $M'$ is used, the new coupling $\lambda' = \lambda+ \frac{3 \lambda^2}{32\pi^2}\log\frac{{M'}^2}{M^2}$ yields the same form for $V_{eff}$.
\end{quote}}

(ii) as $m \to 0$ the effective potential simplifies to

\be
V_{eff}(\varphi_{cl}) = \frac{\lambda}{24}\varphi_{cl}^4\left(1 + \frac{3}{32\pi^2}\lambda \Big[ \log(\frac{\varphi_{cl}^2}{M^2}) - \frac{25}{6}\Big]\right).
\label{Veff-Eq}
\ee
This has a minimum at

\be
\varphi_\star^2 \approx M^2 {\rm e}^{- \frac{32\pi^2}{3\lambda}}.
\ee
However, the minimum is obtained by equating leading and next order (quantum) terms and thus implies that perturbation theory is invalid, and hence the new minimum is likely invalid\footnote{
Sometimes the situation can be improved with the aid of the renormalisation group equations (see section \ref{renormalisation} for more information). For the effective potential this is
$$
\left[ M \frac{\partial}{\partial M} + \beta \frac{\partial}{\partial \lambda} - \gamma \varphi_{cl} \frac{\partial}{\partial \varphi_{cl}}\right] V_{eff}(\varphi_{cl}; M, \lambda) = 0.
$$
Solving this equation and matching it to the perturbative result (recall that first order partial differential equations only give constraints on the solutions) gives
$$
V_{eff}^{\rm RGI} = \frac{\lambda}{24} \,\varphi_{cl}^4\,  \frac{1}{1-\frac{3}{32\pi^2}\lambda \log\varphi_{cl}^2/M^2},
$$
which is 
called the {\it renormalisation group improved} effective action.}.

\boxit{\begin{quote}
Show that the minimum of massless $\varphi^4$ theory in three dimensions is perturbatively stable at one loop order.
\vskip .3 cm
Ans. $\varphi_\star = \frac{3}{4\pi}\sqrt{\frac{\lambda}{2}}$.
\end{quote}} 

\subsubsection{Symmetry Breaking}

Symmetry breaking happens when the ground state of a system does not have the full symmetry of the action. Unfortunately, the terminology is something of a malapropism as the symmetry is not really broken, rather it is hidden. 
Note also that some authors distinguish {\it spontaneously broken symmetry} (a scalar field acquires a vacuum expectation value) from {\it dynamically broken symmetry} (in which symmetry is hidden even when there are no scalar fields). There is a major practical difference between these: spontaneous symmetry breaking can implemented by simply assuming it (this is what is done in the Standard Model), while dynamical symmetry breaking requires a nonperturbative understanding of the system.

Whether symmetry is broken or not, the effective potential must maintain the symmetry of the action. For example, in $\varphi^4$ theory the effective action must be even. The reason is that if it were not true, then infinities in the coefficients of the new terms would require renormalisation via a Lagrangian parameter that does not exist. Thus renormalisable field theories have symmetric effective actions.

So how does symmetry breaking occur?
It is possible for an even function to have two degenerate minima at, say, $\varphi_{cl} = \pm v$. In quantum mechanics any initial state centred on one minimum can tunnel to the other minimum and eigenstates have even or odd parity. Quantum field theoretic (or classical systems in the bulk limit) have infinitely many degrees of freedom, which one can think of as taking infinitely long to tunnel from one local minimum to another. Thus it is possible for an eigenstate to have mixed symmetry. This is what happens when a spin system acquires an induced magnetisation. So it
is possible that the classical field stays in the vicinity of, say, $\varphi_{cl} = +v$, giving rise to spontaneous symmetry breaking.

The computation of the previous section begs the question: is it possible for quantum effects to break a classical symmetry? Although we cannot say if it happened for $\varphi^4$ theory, it {\it is} possible in general. A classic example is broken scalar QED, called the {\it Coleman-Weinberg model}. Ideas like this have also been used to compute bounds on the Higgs mass.

Since most techniques develop the effective potential as a power series in the classical field, the simplest way to ensure reflection symmetry is to expand about the symmetric point $\varphi_{cl}=0$. Of course one could also expand about another point, say $\varphi_{cl} = v$. In this case odd terms would be generated, for example $\Gamma^{(3)} = -i \lambda_0 v$ at tree order. But the sum of all terms will generate an even effective action and renormalising the even terms will lead to complete renormalisation of the odd terms.

\subsection{Goldstone's Theorem}
\label{goldstone}

[This section follows the presentation in chapter 5 of Ref. \cite{amit}.]

We illustrate the power of the functional formalism by sketching a simple derivation of Goldstone's theorem. The theorem asserts that when a classical continuous symmetry is broken a massless particle necessarily exists in the spectrum.

Consider the $O(2)$ linear sigma model. This Lagrangian has an $O(2)$ symmetry under rotations of the field:

\be
\phi' = {\cal R}\phi
\ee
where ${\cal R}$ is a 2x2 rotation matrix and 

\be
\phi = \left(\begin{array}{c}\pi \\ \sigma\end{array}\right).
\ee
 For small rotation angles, $\epsilon$

\be
{\cal R} = 1 + \epsilon \left( \begin{array}{cc} 0 & -1 \\ 1 & 0 \end{array}\right).
\ee

Since the Lagrangian is invariant under this transformation (and it is on fields only), the action, the generating functional, and the effective action are also invariant. Thus

\begin{eqnarray}
Z[J] &=& \int D\phi\, {\rm e}^{ i \int {\cal L}[\phi] + i \int J \phi} \nonumber \\
     &=& \int D\phi'\, {\rm e}^{ i \int {\cal L}[\phi'] + i \int J \phi'} \nonumber \\
     &=& \int D\phi\, {\rm e}^{ i \int {\cal L}[\phi] + i \int J {\cal R}\phi} \nonumber \\
     &\approx& Z[J] + \epsilon \int D\phi\, \int \left(J_\sigma \pi - J_\pi \sigma\right)  {\rm e}^{ i \int {\cal L}[\phi] + i \int J \phi}.
\end{eqnarray}

The term proportional to $\epsilon$ must vanish and we get

\be
\int \left( \frac{\delta \log Z }{\delta J_\pi} J_\sigma  - \frac{\delta \log Z}{\delta J_\sigma} J_\pi\right) = 0.
\ee
The $\log$ is there because we want to normalise. Now make the Legendre transformation (recall $J_i = - \delta \Gamma/\delta \varphi_{cl}^{(i)}$):

\be
\int\left(\pi_{cl} \frac{\delta \Gamma}{\delta \sigma_{cl}} - \sigma_{cl} \frac{\delta \Gamma}{\delta \pi_{cl}} \right) = 0.
\label{WIEq}
\ee
This is a {\it Ward identity} for the sigma model that is true because of its $O(2)$ symmetry, regardless of whether the symmetry is hidden or not.

To get the Goldstone theorem take a derivative with respect to $\pi_{cl}(y)$ to get

\be
\int d^dx \left( \pi_{cl}(x) \frac{\delta^2 \Gamma}{\delta\pi_{cl}(y) \delta \sigma_{cl}(x)} + \delta(x-y) \frac{\delta \Gamma}{\delta \sigma_{cl}(x)} - \sigma_{cl}(x) \frac{\delta^2\Gamma}{\delta \pi_{cl}(y)\delta\pi_{cl}(x)} \right) = 0.
\ee
Finally, assume a hidden symmetry:

\be
\phi_{cl} = \left( \begin{array}{c} \pi_{cl} \\ \sigma_{cl} \end{array} \right) =   \left( \begin{array}{c} 0 \\ v \end{array} \right).
\ee
One says that the symmetry is hidden (or broken) because this matrix element does not obey the $O(2)$ symmetry of the Lagrangian.

Substituting into the last expression gives the desired result

\be
\frac{\delta\Gamma}{\delta \sigma_{cl}(y)} - \int d^d x \, v \frac{\delta^2\Gamma}{\delta \pi_{cl}(y)\delta \pi_{cl}(x)} = 0
\ee
or
\be
v \int d^d x \, \frac{\delta^2 \Gamma}{\delta \pi_{cl}(y)\delta \pi_{cl}(x)} = -J_\sigma(y).
\ee
As the $\sigma$ source goes to zero one must either have that $v\to 0$ (but this is the symmetric vacuum, which we do not consider) or

\be
\lim_{p\to 0} \Gamma_{\pi\pi}(p) = 0.
\ee
We have performed a Fourier transform and adopted a compact notation. 
We will show (see Eq. \ref{InvPropEq} below) that the second derivative of $\Gamma$ is the exact inverse propagator; this implies that the pion 
propagator has a pole at $p^2=0$, which means it is massless.

\boxit{\begin{quote}
[Amit, 5.11] Derive the relation
$$
\Gamma_{\pi\pi}(p) - \Gamma_{\sigma\sigma}(p) = -v \Gamma_{\sigma\pi\pi}(p,0,-p).
$$
\vskip .3 cm
\noindent
Ans: Start with the Ward identity and take derivatives with respect to $\pi_{cl}$ and $\sigma_{cl}$.
\end{quote}}

\subsection{Transition Elements -- Quantum Mechanics}
\label{transition1}

The path integral is an infinite version of the familiar Riemann integral and hence should admit familiar integral theorems such as integration by parts. This can provide powerful nonperturbative information on quantum field theories. Some of this information is explored in this section.

We start by reviewing some freshman calculus. Consider the integral

\be
 I = \int_{-\infty}^\infty dx \, e^{-{1 \over 2} k x^2} 
\ee
and let
$x = y + \Delta f(y)$. One obtains

\be
  I' = \int_{-\infty}^\infty d y \, (1 + \Delta f') \, e^{-{1 \over
2} k (y + \Delta f)^2} = I.
\ee
but $I'(\Delta = 0) = I$ so that the non-zero order terms in $\Delta$ 
must all vanish identically. Thus

\be
 \int_{-\infty}^{\infty} dy \, e^{-{1 \over 2} k y^2} \, (f' - kyf) 
= 0
\ee
for any function $f$. This, at first sight remarkable, equation is 
simply a restatement of integration by parts and so is true for all 
$f$ that diverge more slowly than $e^{{1\over 2}k y^2}$.

Let's consider a more general example.

\be
 I(a,k,V) = \int dx \, e^{-{1 \over 2} k x^2 + V(x) + ax}.
\ee

\noindent
Proceeding as before and using $ \int dx \, f(x) \, e^{ax} = f({d 
\over da}) \, \int dx \, e^{ax}$ gives

\be
\bigl[ f'({d \over da}) - k {d \over da} f({d \over da}) + a f({d
 \over da}) + V'({d \over da}) f({d \over da}) \bigr] I(a;k,V) = 0.
\label{trickEq}
\ee

\noindent
For example if $V = -{\lambda \over 4} x^4$ and $ f = 1$ then

\be
 \lambda I'''(a) + kI'(a) - aI(a) = 0,
\ee

\noindent
which may be solved perturbatively in $\lambda$ for the behaviour of $I(a)$. Of course this equation also lends itself to nonperturbative studies.

Let's apply these ideas to quantum mechanics.  Start with the detailed expression for the path integral of Eq. \ref{SdefEq}

\be
\langle x_f;t_f \vert x_i;t_i \rangle =  \int \prod_{j = 1}^N 
{dx_j \over A} \, e^{{i \over \hbar \epsilon} \sum_{j=0}^N {m \over 2}
(x_{j+1} - x_j)^2 - \epsilon^2  V({x_{j+1} + x_j \over 2})} 
\ee
and make the transformation $x_j \to y_j + \delta f_j$, with $f_j = f(y_j)$:

\be
\langle x_f;t_f \vert x_i;t_i \rangle =
   \int \prod_{j= 1}^N {dy_j \over A} \, (1 + \Delta 
f'_j) \, e^{{i \over \hbar \epsilon}\sum_{j=0}^N [{m \over 2}(y_{j+1} - 
y_j + \Delta(f_{j+1} 
- f_j))^2 - \epsilon^2 V({y_{j+1} + y_j \over 2} + \Delta({f_{j+1} + f_j
 \over 2}))]}
\label{transEq}
\ee
where $x_0 = x_i$, $x_{N+1} = x_f$, $y_0 = y_i + 
\Delta f_i = x_i$, and $y_{N+1} = y_f + \Delta f_f = x_f$. But

\begin{eqnarray}
\sum_{j = 0}^N (y_{j+1} - y_j + \Delta(f_{j+1} - f_j))^2 &=& 
\sum_{j=0}^N (y_{j+1} - y_j)^2 + 2\Delta \sum_{j=0}^N (y_{j+1} - y_j)(f_{j+1} 
- f_j) + \nonumber \\
 \phantom{x} && \qquad 2\Delta(\dot y_i f_i - \dot y_f f_f) + 
 O(\Delta^2,\epsilon^2)
\end{eqnarray}

\noindent
where now $y_0 = y_i$ and $y_{N+1} = y_f$ so that the path integral is 
now in its usual form. The order $\Delta$ term in Eq. \ref{transEq} is

\be
 \int Dy \, e^{{i \over \hbar} S_{cl}(\dot y, y)} \, \bigl[ \sum_{j=
1}^N f'_j + {i \over \hbar} \int_{t_i}^{t_f} (m \dot y \dot f - V'(y) f) 
\, dt + {i m \over \hbar} (\dot y_i f_i - \dot y_f f_f) \bigr] = 0
\ee

\noindent
integrating $\dot y \dot f$ by parts then gives

\be
 \bigl\langle \sum_{j=1}^N f'_j \bigr\rangle = {-i \over \hbar} 
\bigl\langle \int_{t_i}^{t_f} {\delta S \over \delta y(t)} f(y(t)) \, 
dt \bigr \rangle
\ee

This is a remarkable and powerful equation. For example if $f_j =
\delta_{jk}/\epsilon$ ($f(y(t)) = \delta(t - k)$) then

\be
\langle m \ddot x \rangle = -\langle V' \rangle,
\ee
which is the quantum analogue of Newton's equation.

Finally consider $f_j = y_j \delta_{jk} / \epsilon$ or $f(y(t)) = y(t) 
\delta (t-k)$ then

\begin{eqnarray}
\langle 1 \rangle &=& {i \epsilon \over \hbar} \langle
 m \ddot y y(k) + V(y(k)) y(k) \rangle \nonumber \\
\phantom{x} &\sim& {i \over \hbar} \langle m y_k {(y_{k+1} - 2 y_k +
y_{k-1}) \over \epsilon} \rangle \nonumber \\
\phantom{x} &\sim& {i \over \hbar} \langle m y_k \bigl({y_{k+1} -  y_k 
\over \epsilon} - {y_k - y_{k-1} \over \epsilon} \bigr)\rangle \nonumber \\
\phantom{x} &\sim& {i \over \hbar} \langle m {y_{k+1} - y_k \over
 \epsilon}  (y_{k} -  y_{k+1}) \rangle .
\end{eqnarray}

\noindent
Or

\be
 \langle {(y_{k+1} - y_k)^2 \over \epsilon^2} \rangle \sim \langle 
\dot y ^2 \rangle \sim -{\hbar \over i \epsilon} \langle 1 \rangle 
\ee

\noindent
The right hand side goes to infinity as $\epsilon$ goes to zero so that the 
important paths are {\it not} those with a well-defined slope.  In other words, infinitely spiky paths dominate the sum over paths, in keeping with the statement made in Eq. \ref{lipsEq}.

\subsection{Transition Elements -- Quantum Field Theory}
\label{transition2}

The field theoretic analogue of the previous expressions is developed in the same way. Here we choose to work with the continuum notation.
Consider the change of variables $\varphi \to \chi + \epsilon F(\chi)$ in the generating functional of a scalar field theory:

\be
Z[J] = \int D\varphi\, {\rm e}^{iS + i\int J\varphi}.
\ee
\begin{eqnarray}
Z[J] &=& \int D\chi\ \, \vert\frac{\delta\varphi}{\delta \chi}\vert \, {\rm e}^{iS[\chi + \epsilon F] + i\int J\chi + i\epsilon\int J F}\nonumber \\
 &=& \int D\chi\, \vert \delta^d(x-y)(1 + \epsilon \frac{\delta F}{\delta \chi})\vert \, {\rm e}^{iS[\chi] + i\int J\chi} \left( 1+i\epsilon\int \frac{\delta S}{\delta \chi} F + i\ \epsilon\int J F + \ldots\right) \nonumber \\
 &=& \int D\chi\, \left(1 + \epsilon\,\delta^d(0)\, \int \frac{\delta F}{\delta \chi}\right) \, {\rm e}^{iS[\chi] + i\int J\chi} \left( 1+i\epsilon\int \left(\frac{\delta S}{\delta \chi} F +  J F\right)\right).  \\
\end{eqnarray}
The term in the first brackets above is the measure; in discretised form it is 
\be
\prod d \chi_i (1 + \epsilon F_i') \sim \prod d\chi_i + \epsilon \prod d \chi_i \frac{1}{\Delta} \Delta \sum_j F_j' \sim D\chi + D \chi \epsilon \delta(0) \int F',
\ee
where $1/\Delta$  has been written as $\delta(0)$.

As before, the order $\epsilon$ term must evaluate to zero, hence

\be
\int d^dx \left[ \delta^d(0) F'\left(\frac{\delta}{i\delta J(x)}\right) + i\left( \lim_{y\to x} \frac{\delta S}{\delta\varphi}\left(\frac{\delta}{i\delta J(y)}\right) + J(x)\right) F\left(\frac{\delta}{i\delta J(x)}\right) \right] Z[J] = 0.
\label{S1Eq}
\ee
The limit is introduced to remove ambiguity in taking derivatives. 

If $F=1$ one obtains the {\it quantum equations of motion}

\be
\langle (\partial^2 + m_0^2)\chi + V'(\chi)\rangle = \langle J \rangle.
\ee
We will shortly see that this equation, written in terms of the effective action, is equivalent to the Schwinger-Dyson equations.

\boxit{\begin{quote}
Set $V=0$ and $F=\frac{1}{3}\chi^3$ and confirm Eq. \ref{S1Eq}.
\end{quote}}

\subsection{The Schr\"odinger Functional}
\label{schrodinger}

Before moving on to a description of the Schwinger-Dyson equations, we consider one last application of the functional formalism. 

The path integral maintains a prominent place in quantum field theory because it is a compact representation of the quantisation process and because it is manifestly Lorentz invariant. Nevertheless, one can ask where, if anywhere, the Hamiltonian fits into a functional formalism. In fact  all of the familiar language of Hamiltonian quantum mechanics maps over (with some modification) to quantum field theory. Thus one speaks of a Hamiltonian density, the vacuum wavefunctional, eigenstates, variational principles, and so on.

The Hamiltonian for a given theory can be derived from Eq. \ref{psidotEq} since temporal evolution is controlled by $H$. Extending to the quantum field case yields

\be
\Psi[\varphi; t+\epsilon) = \int D\eta \, {\rm e}^{i S[\eta\to\varphi] } \Psi[\eta; t)
\ee
We stress that $\Psi$ is a wavefunctional that depends on a field, $\varphi$ is the final field defined at time $t +\epsilon$; $\eta$ is defined at time $t$. Expanding the action in powers of $\epsilon$ and recalling that $\delta x^2 \sim \epsilon$ gives

\be
i \frac{\partial}{\partial t}\Psi[\varphi] = \int d^{d-1}x \, {\cal H} \Psi[\varphi]
\ee
with
\be
{\cal H} = -\frac{1}{2}\frac{\delta^2}{\delta \varphi^2} +\frac{1}{2}(\nabla \varphi \cdot \nabla \varphi +  m_0^2 \varphi^2) + \frac{\lambda_0}{24} \varphi^4.
\ee

It is tempting to interpret $\langle\varphi|\Psi\rangle$  as the amplitude for the field to realise the specific value $\varphi$, but we should remember that only S-matrix elements carry a physical interpretation in quantum field theory.

 Let's push the analogue further and consider a variational estimate of the vacuum functional. We start with a guess for the free action:

\be
\langle \varphi|\Psi_0\rangle \propto {\rm e}^{-\frac{1}{2}\int \frac{d^{d-1}k}{(2\pi)^{d-1}} \varphi(-k) \omega(k) \varphi(k)}.
\ee
Evaluate 

\be
E_{\rm trial}[\omega] = \frac{\langle \Psi_0 | H_0 = \int d^{d-1}x {\cal H}_0 | \Psi_0\rangle}{\langle \Psi_0|\Psi_0\rangle}
\ee
and recall that 

\be
\langle \Psi | \eta\rangle = \int D\varphi \, \Psi^*[\varphi] \eta[\varphi].
\ee

It is convenient to rewrite $H_0$ in momentum space (and we consider $d=4$):

\be
H_0 = \frac{1}{2}\int \frac{d^3k}{(2\pi)^3}\, \left( \frac{\delta^2}{\delta \varphi_k^2} + (k^2 + m_0^2) \varphi_k^2\right).
\ee
Thus
\be 
E_{\rm trial} = \frac{\frac{1}{2}\int D\varphi\, \int \frac{d^3q}{(2\pi)^3} \left( \omega_q (2\pi)^3\delta^3(0) + (q^2+m_0^2-\omega_q^2)\varphi_q^2\right) {\rm e}^{-\int \varphi_k \omega_k \varphi_{-k}}}{\int D\varphi\, {\rm e}^{-\int \varphi_k \omega_k \varphi_{-k}}}.
\ee
Evaluating the functional integral gives
\be
E_{\rm trial}[\omega] = \frac{1}{4} \delta^3(0) (2\pi)^3\, \int \frac{d^3q}{(2\pi)^3} \left( \omega_q +\frac{q^2+m_0^2}{\omega_q}\right).
\ee
Take the functional derivative with respect to the unknown $\omega$ and solve the equation to obtain
\be
\omega_k = \sqrt{k^2 + m_0^2}.
\ee
Thus the ground state functional for free scalar field theory is given exactly by a Gaussian functional with a kernel specified by the free particle dispersion relation. Finally, substituting gives the familiar ground state energy density:

\be
\frac{E_0}{V} = \frac{1}{2}\int \frac{d^3 q}{(2\pi)^3} \sqrt{q^2+m_0^2}.
\ee

\boxit{\begin{quote}
\noindent
Use the same Ansatz and obtain the equation for the ground state wavefunctional for $\varphi^4$ theory.
\vskip .3 cm
Ans: $\omega_k^2 = k^2 + m_0^2 + \frac{\lambda_0}{4}\int \frac{d^3q}{(2\pi)^3} \frac{1}{\omega_q}$
\end{quote}}

One final remark: Feynman has commented that the variational principle must be useless in quantum field theory because the set of functions probed in any variational Ansatz has measure zero. The problem above indicates that this is too pessimistic: we shall see shortly that the equation for $\omega$ is precisely the same as a truncated Schwinger-Dyson equation for the propagator.

\section{Schwinger-Dyson Equations}
\label{schwingerdyson}

\subsection{The Schwinger-Dyson Master Equation}
\label{schwinger}

In the functional approach, the perturbative evaluation of any n-point function for $\varphi^4$ theory proceeds from the expression

\be
Z[J] = {\rm e}^{iV\left(\frac{\delta}{i \delta J}\right)} \, Z_0[J]
\ee
with $V = \frac{\lambda_0}{24} \varphi^4$. Since $Z_0$ is a  known quadratic form, $Z[J]$, and hence all matrix elements of products of fields, is computable simply by taking repeated derivatives.

Schwinger-Dyson equations are one way (there are many) to organise the diagrams that contribute to 
n-point functions in a quantum field theory. They are popular because they naturally sum infinitely many diagrams and therefore automatically contain nonperturbative information.

The starting point is the simple identity (set $F=1$ in Eq. \ref{S1Eq}):

\begin{equation}
\int D\varphi \left[ \frac{\delta S}{\delta \varphi} + J\right] {\rm e}^{iS + i \int J\varphi} = 0.
\label{SDE-start}
\end{equation}
Extracting the functional derivative using a standard trick (see Eq. \ref{trickEq}) and substituting
the definition of the connected generated functional (Eq. \ref{FdefEq}) gives

\be
{\rm e}^{-iF} \frac{\delta S}{\delta \varphi(x)}\left[ \frac{\delta}{i\delta J}\right ] {\rm e}^{iF} = -J(x)
\label{m0Eq}
\ee
While workable, this equation is problematic because we prefer to work with
irreducible n-point functions, which means eliminating all references to $F$. In general this is quite painful, involving more and more elaborate substitutions as one moves up the sequence of Schwinger-Dyson equations. Fortunately there is a simple trick that eliminates references to $F$ from the start: the previous equation is equivalent to

\begin{equation}
\frac{\delta S}{\delta \varphi(x)} \left[ \frac{\delta}{i\delta J} + \frac{\delta F}{\delta J}\right] \cdot 1 = - J(x)
\label{m1Eq}
\end{equation}
The `1' on the left hand side indicates that one should take derivatives of all quantities to the right of a derivative and that one stops upon taking the derivative of unity. Thus, for example, the functional derivative appearing in the argument of $\delta S/\delta \varphi$ only acts on $\delta F/\delta J$.

\boxit{\begin{quote}
\noindent
Obtain Eq \ref{m1Eq} from Eq. \ref{m0Eq}.
\vskip .3 cm
\noindent
Ans: work in perturbation theory or follow the strategy employed in proving the Baker-Campbell-Hausdorff formula.
\end{quote}}

Now make the Legendre transformation, using the chain rule to replace the derivative with respect to the source with:

\be
\frac{\delta}{i\delta J} = \int d^d z\,\frac{\delta \varphi_{cl}(z)}{i\delta J} \frac{\delta}{\delta \varphi_{cl}(z)}
\ee
then replace 

\be
\frac{\delta \varphi_{cl}(z)}{\delta J} = \frac{\delta^2 F}{\delta J \delta J(z)}.
\label{m2Eq}
\ee

We are almost there, we only need obtain an expression for the second derivative of $F$ in terms of the effective action. Consider, therefore,

\begin{equation}
\delta(x-z) = \frac{\delta \varphi_{cl}(x)}{\delta \varphi_{cl}(z)} = \int d^d y \frac{\delta^2 F}{\delta J(x) \delta J(y)} \frac{\delta J(y)}{\delta \varphi_{cl}(z)} = -\int d^d y \frac{\delta^2 F}{\delta J(x) \delta J(y)} \frac{\delta^2 \Gamma}{\delta \varphi_{cl}(z) \delta \varphi_{cl}(y)}
\end{equation}

Thus

\begin{equation}
\frac{\delta^2 \Gamma}{\delta \varphi_{cl}(x) \delta \varphi_{cl}(z)}= -\left(\frac{\delta^2 F}{\delta J(x) \delta J(z)}\right)^{-1}.
\label{FinvEq}
\end{equation}
This is the relation we need to complete the derivation of the Schwinger-Dyson master equation. Before writing this we first note that setting the source equal to  zero implies that the vacuum expectation value of the field is also zero

\be
\varphi_{cl}|_{J=0} = 0.
\ee
(We shall discuss a more general situation below).
Finally, Eq. \ref{FinvEq} implies that

\begin{equation}
\frac{\delta^2 \Gamma}{\delta \varphi_{cl}(x) \delta \varphi_{cl}(z)}|_{\varphi_{cl}=0} = i \Delta^{-1}(x-z).
\label{InvPropEq}
\end{equation}
{\it The irreducible two point function is the inverse of the exact propagator.}

We now continue with the derivation of the Schwinger-Dyson master equation; substituting Eq. \ref{FinvEq} into Eq. \ref{m1Eq} along with Eqs. \ref{m2Eq} and \ref{FEq}  gives

\begin{equation}
\frac{\delta S}{\delta \varphi}\left[ \varphi_{cl} + i \int \left(\frac{\delta^2 \Gamma}{\delta\varphi_{cl}\delta\varphi_{cl}}\right)_{.z}^{-1} \frac{\delta}{\delta \varphi_{cl}(z)}\right] \cdot 1 = \frac{\delta\Gamma}{\delta \varphi_{cl}}
\label{m3Eq}
\end{equation}
This is the master equation for generating all Schwinger-Dyson equations for a scalar field theory in terms of irreducible n-point functions.

Let us make this more concrete by considering $\varphi^4$ theory. Recall that (Eq. \ref{eomEq})

\begin{equation}
\frac{\delta S}{\delta \varphi} = -(\partial^2 + m_0^2) \varphi - \frac{\lambda_0}{6} \varphi^3.
\end{equation}
Thus the master equation becomes

\begin{equation}
-(\partial^2 + m_0^2) \varphi_{cl}(x) - \frac{\lambda_0}{6}\left(\varphi_{cl}(x) + \Delta_{xz} \frac{\delta}{\delta \varphi_{cl}(z)}\right)^3 \cdot 1 = \frac{\delta \Gamma}{\delta \varphi_{cl}(x)}
\label{dD}
\end{equation}

Simplify using 

\begin{equation}
\frac{\delta}{\delta\varphi(z)}\Delta_{xy} = i\frac{\delta}{\delta \varphi(z)}\left(\frac{\delta^2\Gamma}{\delta \varphi \delta\varphi}\right)^{-1}_{xy} = \Delta_{xa}(i\Gamma_{azb})\Delta_{by}
\label{dDEq}
\end{equation}
We have defined the three point function

\be
\Gamma_{abc} = \frac{\delta^3\Gamma}{\delta\varphi(a)\delta\varphi(b)\delta\varphi(c)}.
\ee

Henceforth all fields will be classical fields, and we drop the ${cl}$ subscript.

\boxit{\begin{quote}
\noindent
 prove Eq. \ref{dDEq}
\vskip .3 cm
\noindent
Ans: consider this as a matrix equation and use $\frac{d}{dx} (M M^{-1}) = 0$.
\end{quote}}

The final form of the master equation for $\varphi^4$ theory is thus:

\begin{equation}
-(\partial^2 + m_0^2)\varphi_x - \frac{\lambda_0}{6}\left( \varphi^3_x +3 \Delta_{xx}\varphi_x + 
\Delta_{xa}\Delta_{xb}\Delta_{xc} i\Gamma_{abc}\right) = \frac{\delta \Gamma}{\delta \varphi_x}
\label{MEq}
\end{equation} 

\subsection{Schwinger-Dyson Diagrammatics}
\label{schwinger2}

Let us consider setting the source equal to zero in the master equation for $\varphi^4$ theory, Eq. \ref{MEq}. The defining equation \ref{J-eqn} then implies that the left hand side is zero. If we take $\varphi_{cl}|_{J=0} = 0$ as discussed above, then this implies the last term on the left hand side is zero, which implies that the three point function is zero

\be
\Gamma_{abc} = 0.
\ee
This is sensible: $\varphi^4$ theory is invariant under parity reflections and there should be no odd n-point functions if one expands about $\varphi_{cl}=0$.

The Schwinger-Dyson equation for the propagator follows by taking the derivative of the master equation with respect to the classical field:

\begin{equation}
\frac{\delta^2 \Gamma}{\delta \varphi_x\delta\varphi_y} = -(\partial^2_x + m_0^2) \delta_{xy} - \frac{\lambda_0}{2} \varphi_x^2 \delta_{xy} - \frac{\lambda_0}{2} \frac{\delta}{\delta \varphi_y}(\Delta_{xx}\varphi_x) - \frac{\lambda_0}{6} \frac{\delta}{\delta\varphi_y}(\Delta_{xa}\Delta_{xb}\Delta_{xc}i\Gamma_{abc}).
\end{equation}
Notice that repeated $x$ indices are not summed in the above; in fact the delta functions in $x$ and $y$ are important for routing momenta. However, the letters at the beginning of the alphabet {\it are} summed over!  Taking the derivatives with the aid of Eq. \ref{dDEq}, setting the source equal to zero, and recognising that three terms are in fact identical, gives

\begin{equation}
\frac{\delta^2 \Gamma}{\delta \varphi_x\delta\varphi_y}\vert_{J=0} = -(\partial^2_x + m_0^2) \delta_{xy} - \frac{\lambda_0}{2} \Delta_{xx}\delta_{xy} - \frac{\lambda_0}{2} \Delta_{xb}\Delta_{xc}\Delta_{xd}i\Gamma_{abc}\Delta_{ea}i\Gamma_{dye} - \frac{\lambda_0}{6} \Delta_{xa}\Delta_{xb}\Delta_{xc}i\Gamma_{abcy}
\label{2ptEq}
\end{equation}

\boxit{\begin{quote}
Exercise: verify this equation to first order in perturbation theory.
\end{quote}}

\begin{figure}[ht]
\includegraphics[width=9cm]{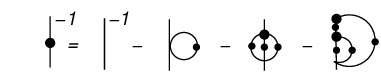}
\caption{The propagator Schwinger-Dyson Equation}
\label{phi-4-prop-SDE-fig}
\end{figure}

It is convenient to rewrite this equation in  terms of diagrams. While it is possible to develop these in terms of spacetime coordinates, it is useful to work in momentum space to take advantage of the translational invariance of the system. 
First we recall that the left hand side is the inverse full propagator; similarly the first term on the right hand side is the inverse of the lowest (tree) order propagator given in Eq. \ref{deltaEq}. The second term contains a full propagator that runs from $x$ to $x$ and a `stub' that runs from $x$ to $y$. Similarly, the last term involves three full propagators that run from $x$ to a four-point vertex, whose last leg connects to the point $y$. Finally, we can write Eq. \ref{2ptEq} as shown in Fig. \ref{phi-4-prop-SDE-fig}.

Following tradition, full propagators and vertices are represented with dots.

We note the following:

(i) diagram topology is specified by following indices in the Schwinger-Dyson equation

(ii) factors of i, 1/2, -1, etc are absorbed into the definition of the diagrams {\it except} as indicated in Fig. \ref{phi-4-prop-SDE-fig}. This is because the perturbative Feynman rules are sufficient to determine all of these factors uniquely. The explicit minus signs then make everything work out

(iii) the master equation implies that there must be exactly one bare vertex in every Schwinger-Dyson equation

(iv) the master equation restricts the form of possible diagrams that appear in Schwinger-Dyson equations

\boxit{\begin{quote}
\noindent
show that \includegraphics[width=1.5cm]{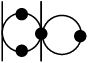} cannot occur in any Schwinger-Dyson equation. 
\vskip .3 cm
\noindent
Ans: use the diagrammatic language below, the starting diagrams, and the fact that there is no `joining' operation.
\end{quote}}

\begin{figure}[ht]
\includegraphics[width=11cm]{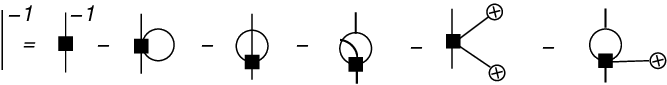}
\caption{Full Propagator with Source Terms}
\label{phi-4-prop-3-SDE}
\end{figure}

Although we have considerably streamlined the process of generating Schwinger-Dyson equations, taking derivatives gets tedious rapidly. Fortunately, the simplicity of the formalism permits an entirely diagrammatic approach. First we introduce a classical field as $\varphi_{cl} = \ $
\includegraphics[width=0.3cm,angle=90]{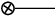}. The master equation can then be written diagrammatically.  One functional derivative yields the full propagator with source terms present, illustrated in Fig. \ref{phi-4-prop-3-SDE}. The notation has been simplified further  by noting that all vertices are full except one, we therefore choose to label the bare four point vertex with a square. All internal lines are also full propagators. Remember that the external lines are stubs (do not propagate).

\begin{figure}[ht]
\includegraphics[width=13cm]{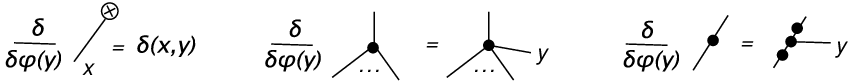}
\caption{Rules for Functional Derivatives.}
\label{derivative-rules}
\end{figure}

One can continue taking derivatives and translating the results into diagrammatic form, but it is easier and faster to work directly in the diagrammatic approach. The required elements are shown in Fig. \ref{derivative-rules}.

\begin{figure}[ht]
\includegraphics[width=10cm]{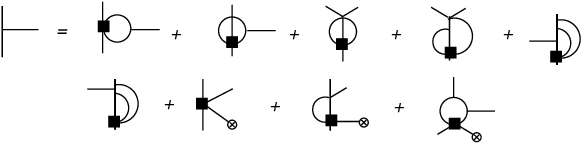}
\caption{Full Three-point Function with Source Terms.}
\label{phi-4-3-pt}
\end{figure}

Applying the rules to Fig. \ref{phi-4-prop-3-SDE} yields the full equation for the three-point function given in Fig. \ref{phi-4-3-pt}.

%\begin{figure}[ht]
%\includegraphics[width=8cm]{master-Eq.eps}
%\caption{master Eq}
%\label{phi-4-prop-SDE}
%\end{figure}

\begin{figure}[ht]
\includegraphics[width=5cm]{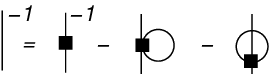}
\caption{The Full Propagator}
\label{phi-4-prop-SDE}
\end{figure}

\begin{figure}[ht]
\includegraphics[width=8cm]{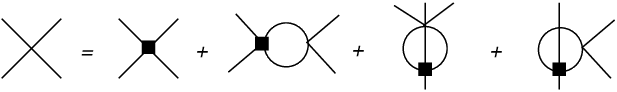}
\caption{The Four Point Function}
\label{phi-4-4-pt-fig}
\end{figure}

\begin{figure}[ht]
\includegraphics[width=9cm]{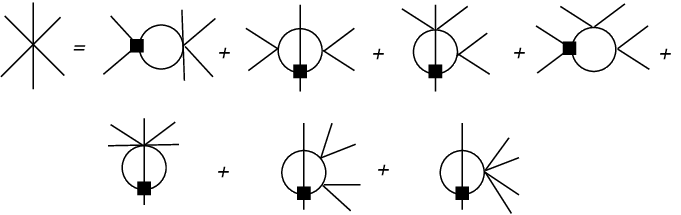}
\caption{The Six Point Function}
\label{phi-4-6-pt-fig}
\end{figure}

With the diagrammatic formalism one can proceed to quite high order. Setting the classical field to zero reduces the number of diagrams dramatically -- we display the first three Schwinger-Dyson equations in Figs. \ref{phi-4-prop-SDE}, \ref{phi-4-4-pt-fig}, \ref{phi-4-6-pt-fig}. It is evident that solving these equations -- the first three of an infinite set -- is a daunting task.
Furthermore, there is no obvious small parameter which one can use to organise the series of Schwinger-Dyson equations. We examine these issues in sections \ref{renormalisation} and \ref{numerical}.

%\begin{figure}[ht]
%\includegraphics[width=8cm]{phi-4-4-2-pt.eps}
%\caption{phi-4-4-2-pt}
%\label{phi-4-prop-SDE}
%\end{figure}

\subsection{Several Fields -- Alternate Schwinger-Dyson Equations}
\label{several}

Schwinger-Dyson equations provide a convenient methodology to sum Feynman diagrams. Of course, one need not sum diagrams in this way. One could, for example, sum all leading infrared divergent diagrams\footnote{This is what is done in the famous resolution of the infrared divergence problem in the degenerate electron gas.}. In fact, even Schwinger-Dyson equations do not admit a unique form. We illustrate this here with a simple two-field bosonic model.

Consider two scalar fields, $A$ and $\varphi$, interacting according to the Lagrangian

\be
{\cal L} = \frac{1}{2}\partial_\mu A \partial^\mu A - \frac{1}{2}\mu^2 A^2 + \frac{1}{2}\partial_\mu\varphi \partial^\mu \varphi - \frac{1}{2} m^2 \varphi^2 - \frac{1}{2} g A\varphi^2.
\ee
The field equations are
\begin{eqnarray}
\frac{\delta S}{\delta A} &=& -(\partial^2 + \mu^2) A - \frac{1}{2} g \varphi^2 \\
\frac{\delta S}{\delta \varphi} &=& -(\partial^2 + m^2) \varphi - g A \varphi .
\end{eqnarray}

Deriving the master equation requires a slight generalisation of the previous method to deal with many fields. Consider, therefore, a set of fields labelled $\varphi_i(x)$. All applications of the chain rule (etc) must sum over the index $i$. Let us  change notation slightly 
\be
\varphi_i(x) \to \varphi(i,x) \to \varphi(X)
\ee
where the variable $X$ is now understood to represent a spacetime coordinate and a field label. In this way all of the previous expressions remain valid, with the understanding that all coordinates are now `generalised coordinates'. As a final bit of notation, if a specific field is specified then I will denote that by, eg,  $X \to \varphi_x$. 

\begin{figure}[ht]
\includegraphics[width=9cm]{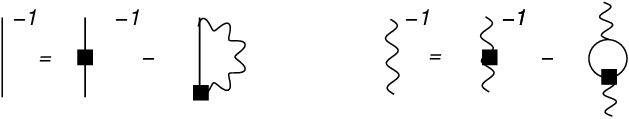}
\caption{Phion and Aon Propagators}
\label{2comp}
\end{figure}

The master equations are
\begin{eqnarray}
\frac{\delta \Gamma}{\delta A(x)} &=& -(\partial_x^2 + \mu^2) A(x) - \frac{1}{2}g \varphi_x^2 - \frac{1}{2} g \Delta_{\varphi_x\varphi_x} \\
\frac{\delta \Gamma}{\delta \varphi(x)} &=& -(\partial_x^2 + \mu^2) \varphi(x) - g A_x \varphi_x  - g \Delta_{A_x\varphi_x}.
\end{eqnarray}
One more derivative yields the propagator equations shown in Fig. \ref{2comp}. The $\varphi$ field is represented as a solid line, while $A$ is denoted with a wiggly line.

\begin{figure}[ht]
\includegraphics[width=9cm]{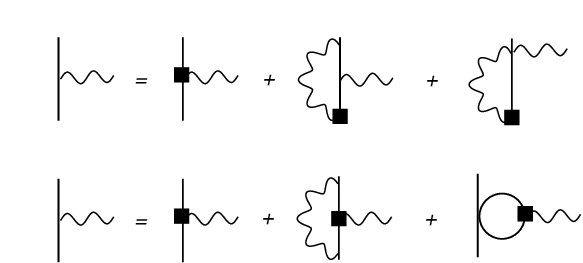}
\caption{Two Vertex Equations in $A\varphi^2$ Theory.}
\label{2compVertex}
\end{figure}

There are several ways to obtain the vertex equation (which is the point of this section). For example one can form

\be
\frac{\delta}{\delta A_z} \frac{\delta^2 \Gamma}{\delta \varphi_x\delta \varphi_y}
\ee
or
\be
\frac{\delta}{\delta\varphi_x}\frac{\delta}{\delta\varphi_y} \frac{\delta\Gamma}{\delta A_z}.
\ee
Doing so yields distinct equations for the full vertex, illustrated in Fig. \ref{2compVertex}. 
We have simplified by assuming that there is not mixed propagator, $\Delta_{\varphi_x A_y}|_{J=0}$, which follows if there is no three-phion vertex. Notice that one cannot conclude that the diagram involving the $\varphi^4$ vertex is equal to that with the $A^2\varphi^2$ vertex because the bare $A\varphi^2$ vertex in the other graph is in a different location in the two versions of the equation. Finally, this discussion carries over in precisely the same form to QED and, with extra diagrams, to QCD.

\boxit{\begin{quote}
Derive these results.
\end{quote}}

\subsection{Some Simple Examples}
\label{some}

\subsubsection{$\varphi^4$ Gap Equation}

Let's rewrite the $\varphi^4$ theory Schwinger-Dyson equation for the two-point function using Feynman rules and representations for the full propagator and vertex:

\be
\Delta_{p} = \frac{i}{F(p^2)}
\ee

\be
\Gamma_4 = \Gamma(\ell_1,\ell_2,\ell_3,\ell_4) \, (2\pi)^d \delta(\ell_1+\ell_2+\ell_3+\ell_4).
\ee

We obtain

\begin{eqnarray}
\frac{F(p^2)}{i} &=& \frac{p^2 -m_0^2}{i} + i \frac{\lambda_0}{2} \int \frac{d^dq}{(2\pi)^d} \frac{i}{F(q^2)} +  \nonumber\\
&& i \frac{\lambda_0}{6}\int \frac{d^d\ell_1}{(2\pi)^d}\, \frac{d^d\ell_2}{(2\pi)^d} \, \frac{d^d \ell_3}{(2\pi)^d} \frac{i}{F(\ell_1^2)} \frac{i}{F(\ell_2^2)} \frac{i}{F(\ell_3^2)} \, \Gamma(\ell_1,\ell_2,\ell_3,-p) \nonumber \\
&& \cdot (2\pi)^d \delta(\ell_1+\ell_2+\ell_3-p).
\label{F2-Eq}
\end{eqnarray}

We see some immediate concerns:

(i) if $F \sim p^2$ and $\Gamma \sim  const$ then the integrals diverge if
$d>1$.

(ii) we need an expression for $\Gamma$. Approximating this as the bare vertex, $\Gamma = -i \lambda_0$, is typical. This is called the {\it rainbow-ladder approximation}.

(iii) we need to evaluate some pretty nasty integrals.

(iv) we need to solve a nonlinear integral equation.

These add up to some pretty hefty demands on our computational and analytic tool bags!

Let's truncate heavily by ignoring the last term. Set $d=4$ and simplify further by performing the Wick rotation to Euclidean space, 

\be
q_0 \to i q_4
\ee
and call $q_E^2 = q_1^2 + q_2^2 + q_3^2 + q_4^2$.
Thus

\be
F(-p_E^2) = -p_E^2 - m_0^2 + \frac{\lambda_0}{2} \int \frac{d^4 q_E}{(2\pi)^4} \frac{1}{F(-q_E^2)}
\ee
Notice that the integral is independent of $p$ hence $F = -p_E^2 + const$. Cut the integral off at $\Lambda$ and call it $-\alpha$ to obtain the consistency condition:

\be
\alpha = 2\pi^2 \int^\Lambda \frac{q_E^3 dq_E}{(2\pi)^4} \frac{1}{q_E^2 + m^2 + \lambda_0 \alpha/2} \qquad (d=4)
\ee

Evaluating the integral for large $\Lambda$ yields a transcendental equation for $\alpha$:

\be
\frac{\lambda_0}{32\pi^2} = \frac{y}{1+y\log y}
\ee
with
\be
y = \frac{\lambda_0 \alpha}{2\Lambda^2}.
\ee
For small $\lambda_0$ on obtains
\be
m^2 = m_0^2 + \lambda_0\frac{\Lambda^2}{32 \pi^2}.
\ee
If $\lambda_0 >32 \pi^2$ then the equation admits no real solution (it has infinitely many complex solutions), implying tachyonic or decaying phions. Presumably this is an indication of a truncation problem or something deeper\footnote{It is believed that $\varphi^4$ theory is trivial in four dimensions.}.

In three dimensions 

\be
m^2 = m_0^2 + \lambda_0 \frac{\Lambda}{4\pi^2} + {\cal O}(\sqrt{\lambda_0^3 \Lambda}).
\ee
Recall that the coupling has units of mass in three dimensions. 

In both cases the phion mass moves to the ultraviolet cutoff, $\Lambda$. This is generally an undesirable situation since one must then cancel two large numbers (the bare mass and the cutoff) very precisely to yield a relatively small physical mass. This is called a {\it fine tuning problem}. Although this example is somewhat academic, the problem is very relevant to current quantum field theory since the Higgs mass requires fine tuning in the Standard Model.

Finally, we make contact with the exercise in section \ref{schrodinger}, in which you were asked to make a functional variational estimate of the ground state energy of $\varphi^4$ theory. The result was a wavefunctional of the form $\langle \varphi|\Psi\rangle \sim \exp(-1/2 \int \varphi \omega \varphi)$ with
\be
\omega_k^2 = k^2 + m_0^2 + \frac{\lambda_0}{4}\int \frac{d^3q}{(2\pi)^3} \frac{1}{\omega_q}.
\ee
We seek the relationship of this result to the Schwinger-Dyson formalism. Since the functional approach is in three spatial dimensions we return to Eq. \ref{F2-Eq}, set $d=4$, neglect the second term, and perform the $q_0$ integral.  To do this note that the integral is independent of momentum and therefore $F(p^2) = p_0^2 - \vec p^2 + const$. Take this into account by setting $F_p = p_0^2 - \omega_p$. Performing the $q_0$ integral then gives

\be
p_0^2 - \omega_p  = p_0^2 - \vec p^2 - m_0^2 - \frac{\lambda_0}{4} \int \frac{d^3 q}{(2\pi)^3} \frac{1}{\omega_q},
\ee
which agrees with the exercise. We see that the functional variational approach yields results that agree with the Schwinger-Dyson equations when truncated at the Hamiltonian level.

\subsubsection{Fermion Contact Model}

Consider a model defined by the Hamiltonian

\be
H = \int d^3x\, \bar \psi ( -i \vec \gamma\cdot \nabla + m)\psi + \frac{1}{2}\int d^3x d^3y\, \bar \psi(y) \gamma_0 T^a \psi V(x-y) \bar\psi\gamma_0 T^a\psi(x)
\ee
where
\be
V(r) = \frac{\lambda}{\Lambda^2}\delta(r_0) \delta(\vec r).
\ee
This model represents relativistic fermions interacting via an instantaneous contact potential. The fermion currents in the interaction are the temporal components of a vector current $\bar \psi \gamma_\mu \psi$ with the addition of a flavour (or colour) structure denoted by the matrix $T^a$.
The coupling $\lambda$ is dimensionless, thus two additional powers (denoted by $\Lambda$) are required to define the interaction. Since this model is not renormalisable we take the scale $\Lambda$ to also be the ultraviolet cutoff of the theory. Finally note that the Lagrangian is invariant under chiral rotations when $m=0$.

\begin{figure}[ht]
\includegraphics[width=5cm]{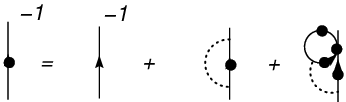}
\caption{Schwinger-Dyson equation for the propagator in the fermion contact model.}
\label{SDE-V}
\end{figure}

The full propagator is given by the Schwinger-Dyson equation shown in Fig. \ref{SDE-V} (this figure does not show the minus signs we have previously made explicit). Notice that the interaction is not dressed in this equation, as one might expect since the potential is a fixed function and not a dynamical quantity. 

The full fermion propagator depends on two scalar functions, hence we employ the Ansatz

\be
S(k) = \frac{i}{A(p^2)\rlap{p}/ - B(p^2)}.
\ee
We truncate by neglecting the four-fermion interaction in the last diagram of Fig. \ref{SDE-V}. 
Substituting gives
\be
\frac{A\rlap{p}/ -B}{i} = \frac{\rlap{p}/-m}{i} - {\rm tr}(T^aT^a) \int \frac{d^4q}{(2\pi)^4} \gamma_0 \frac{i}{A\rlap{q}/ -B} \gamma_0 V(p-q)
\ee
Substitute for $V$ and evaluate the Dirac trace of both sides to obtain an equation for $B$ and multiply by $\rlap{p}/$ and take the trace to obtain an equation for $A$. The result is

\begin{eqnarray}
A(p^2) p^2 &=& p^2  - i\frac{\lambda}{\Lambda^2} C_F \int \frac{d^4 q}{(2\pi)^4} \frac{A(q^2) p\cdot q}{A^2q^2 - B^2} \\
B(p^2) &=& m  + i\frac{\lambda}{\Lambda^2} C_F \int \frac{d^4 q}{(2\pi)^4} \frac{B(q^2)}{A^2q^2 - B^2} 
\label{contact-model-B}
\end{eqnarray}
Notice that the integral in the equation for $A$ is zero hence

\be
A=1
\ee
Furthermore, $B$ must be a constant, and the Euclidean space equation for $B$ is

\begin{equation}
B = m + \frac{\lambda}{\Lambda^2} C_F B \int^\Lambda \frac{d^4 q_E}{(2\pi)^4} \frac{1}{q_E^2 + B^2}.
\end{equation}
Although this equation is quite similar to the $\varphi^4$ case studied above, the extra factor of $B$ in the numerator makes an important difference. Setting $m=0$, dividing by $B$, and evaluating the integral yields

\be
1 = \frac{\lambda C_F}{16 \pi^2} \left( 1 + x^2 \log \frac{x^2}{1+x^2}\right)
\ee
where $x = B/\Lambda$. The term in brackets is bounded by zero and one, hence a nontrivial solution does not exist unless

\be
\lambda > \frac{16 \pi^2}{C_F}.
\ee
Below this value $B=0$. 

We see that there is a critical coupling above which a fermion mass is generated. Thus chiral symmetry is spontaneously broken in this model. This is a nontrivial result that is impossible to obtain in perturbation theory.

Let's return to Eq. \ref{contact-model-B} and perform the integral over $q_0$ using the appropriate contour and Cauchy's theorem. The result is

\be
B = \frac{\lambda C_F}{4\pi^2 \Lambda^2} \int^{\Lambda_3}q^2 dq\,  \frac{B}{\sqrt{q^2+B^2}}.
\ee
We have regulated the integral by cutting it off at a scale $\Lambda_3$. Notice that this amounts to a different regularisation scheme from that adopted in the Euclidean approach. Does this make a difference?

Dividing by $B$ and doing the integral gives

\be
1 = \frac{\lambda C_F}{8 \pi^2} \left( y \sqrt{y^2+x^2} + \frac{1}{2}x^2 \log \frac{x^2}{(y+\sqrt{y^2+x^2})^2} \right)
\ee
where $y = \Lambda_3/\Lambda$.
The term in brackets is bounded by zero and $y^2$ so no solution exists unless

\be
\lambda > \frac{8 \pi^2}{y^2 C_F}.
\ee
We see that selecting $y=1/\sqrt{2}$ yields the same critical coupling as the Euclidean case. While this is reassuring, it is somewhat distressing that the specific dependence of $B$ on the coupling is not the same in the two cases: physics should be independent of the regulator scheme chosen.  The problem here is that the model is not renormalisable and hence residual regulator scheme dependence exists.

\subsubsection{Fermion Contact Model -- Hamiltonian Approach}

We have seen in section \ref{schrodinger} that it is possible to apply functional methods directly to the Hamiltonian. In that section it was shown that the equation for the kernel of a Gaussian wavefunctional Ansatz is equivalent to the Schwinger-Dyson equation for the propagator when truncated at the `Hamiltonian' level (ie, interaction terms are not iterated). The same observation applies to the contact fermion model considered here. Demonstrating this in the functional approach require a generalisation of the method to {\it Grassmann fields}, which are fields with properties appropriate for anticommuting quantum quantities. Since we have not discussed this (and will not), the equivalence to a third approach will be shown here.

Consider a {\it Bogoliubov transformation} which is a canonical transformation in the particle basis:

\begin{eqnarray}
B_k &=& \cos \frac{\phi}{2} \, b_k - \sigma \cdot \hat k \, \sin \frac{\phi}{2}\, d^\dagger_k \nonumber \\
D_k &=& \cos \frac{\phi}{2}\, d_k + \sigma \cdot \hat k \, \sin \frac{\phi}{2}\, b^\dagger_k
\end{eqnarray}

\boxit{\begin{quote}
Show that the Bogoliubov transformation is canonical.
\end{quote}}

\noindent
The function $\phi(k)$ is called the {\it Bogoliubov angle} and it parameterises the structure of the vacuum (recall that particle creation and annihilation operators are defined with respect to an implicit vacuum). To make progress one only need the field contractions

\begin{eqnarray}
\langle \psi_\alpha(x)\psi^\dagger_\beta\psi(y)\rangle &=& \int \frac{d^3k}{(2\pi)^3} [\Lambda_+(k)]_{\alpha\beta} \, {\rm e}^{i\vec k \cdot (\vec x - \vec y)} \nonumber \\
\langle \psi^\dagger_\alpha(x)\psi_\beta\psi(y)\rangle &=& \int \frac{d^3k}{(2\pi)^3} [\Lambda_-(k)]_{\beta\alpha} \, {\rm e}^{-i\vec k \cdot (\vec x - \vec y)}
\end{eqnarray}
with
\be
\Lambda_{\pm} = \frac{1}{2}( 1 \pm \sin \phi \beta \pm \cos \phi \vec \alpha \cdot \hat k).
\ee
Here  $\beta = \gamma^0$ and $\alpha^i = \gamma^0 \gamma^i$.

With these rules one can compute the expectation value of the Hamiltonian with respect to the unknown vacuum. The result is

\begin{eqnarray}
\langle H \rangle &=& -2 N_c \int^\Lambda \frac{d^3 k}{(2\pi)^3}\left(s(k) m + c(k)k\right) \nonumber \\
&&  +
\frac{\lambda}{2\Lambda^2} C_F N_c \int^\Lambda \frac{d^3k}{(2\pi)^3}\, \frac{d^3q}{(2\pi)^3}\, \left( 1 - s(k)s(q) -c(k) c(q) \hat k \cdot \hat q\right)
\end{eqnarray}
with $s(k) = \sin \phi(k)$, $c(k) = \cos \phi(k)$. Then the vacuum energy is minimised 

\be
\frac{\delta}{\delta \phi}\langle H\rangle = 0
\ee
giving
\be
B = m + \lambda \frac{C_F}{4\pi^2\Lambda^2} \int^\Lambda q^2 dq\, \frac{B}{\sqrt{q^2+B^2}}
\ee
where $B = k s(k)/c(k)$.
This is precisely the second form of the gap equation obtained in the previous section. Thus minimising the vacuum expectation value of the Hamiltonian is equivalent to solving the Schwinger-Dyson equations when truncated at rainbow-ladder order.

\subsubsection{Ladder QED}

[This section follows the description in Ref. \cite{SQ}.]

\begin{figure}[ht]
\includegraphics[width=5cm]{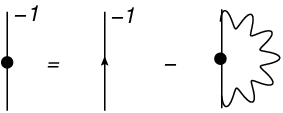}
\caption{Gap Equation for QED in  the rainbow-ladder approximation.}
\label{qed-simple-gap}
\end{figure}

As a final example, we consider quantum electrodynamics in four dimensions in rainbow ladder approximation. The Lagrangian is

\begin{equation}
{\cal L} = -\frac{1}{4} F^{\mu\nu}F_{\mu\nu} -\frac{1}{2\xi}(\partial\cdot A)^2 +\bar \psi(i\partial\cdot\gamma -m)\psi -e \bar \psi A\cdot\gamma\psi.
\end{equation}
The parameter $\xi$ is the (covariant) gauge fixing parameter. 
The Schwinger-Dyson equation for the electron propagator is then given by Fig. \ref{qed-simple-gap}. The Schwinger-Dyson equations have been truncated by employing the bare photon propagator given by
\begin{equation}
D_{\mu\nu} = \frac{-i}{k^2+i\epsilon} \left( g_{\mu\nu} - (1-\xi) \frac{k_\mu k_\nu}{k^2}\right).
\end{equation}
The gauge parameter is often taken to be zero since in this gauge $\xi$ itself is not renormalised (ie, it stays at zero). This is called {\it Landau gauge}.

Following the same procedure as in the previous examples we can derive that $A=1$ in Landau gauge, and
hence

\begin{equation}
B(p^2) = m -i e^2 \int\frac{d^4q}{(2\pi)^4} \frac{(3+0) B(q^2)}{(q^2 - B^2)(p-q)^2}
\label{qed-B-2}
\end{equation}
(the zero above comes from $\xi$). 

Remarkably, the angular integral has a very simple form:

\be
\int d\Omega_4 \frac{1}{(p-q)^2} = 2\pi^2 \left[ \frac{1}{q^2}\theta(q>p) + \frac{1}{p^2}\theta(p>q)\right].
\ee
Substituting, setting $m=0$, Wick rotating, and regulating with a cutoff yields

\begin{equation}
B(-p_E^2) = \frac{3e^2}{8\pi^2}\left[ \frac{1}{p_E^2} \int_0^{p_E} dq q^3 \frac{B(q)}{q^2+B^2} + \int_{p_E}^{\Lambda} dq q 
\frac{B(q)}{q^2+B^2}\right]
\label{B-int-eqn}
\end{equation}
This is a relatively simple integral equation to solve numerically since there are no angular integrals to evaluate and the $p_E$-dependence is easy to implement. However, the equation can solved analytically by converting it into a nonlinear differential equation.

Taking the derivative with respect to $p_E^2$ gives

\be
p_E^4 B' = - \frac{3e^2}{8\pi^2} \int_0^{p_E} dq q^3 \frac{B}{q^2+B^2}
\ee
hence
\begin{equation}
(p_E^4 B')' = -\frac{3 e^3}{16\pi^2} p_E^2 \frac{B}{p_E^2+B^2}
\label{B-deriv-eqn}
\end{equation}

This nonlinear differential equation is equivalent to the integral equation once we have specified the boundary conditions that are implicit in the integral. To obtain them, 
substitute Eq \ref{B-deriv-eqn} into Eq \ref{B-int-eqn} and integrate. The result is

\be
(p^4 B')|_{p=0} = 0
\ee
and
\be
(B + p^2 B')|_{p=\Lambda} = 0
\ee

We proceed by noting that Eq \ref{B-deriv-eqn} is homogeneous for large momenta and hence has the simple power law solution

\be
B \to p^{-1 \pm \sqrt{1 - 3e^2/(4\pi^2)}}
\ee
This satisfies the first boundary condition, but can only satisfy the second if the power is imaginary. 

\boxit{\begin{quote}
\noindent
show this. 
\vskip .3 cm
\noindent
Ans: if the power is imaginary, $B$ is  oscillatory and hence it can be arranged that $(B+p^2B') = 0$ at $p=\Lambda$.
\end{quote}}

Thus a nontrivial gap function is generated (in rainbow ladder approximation in Landau gauge) if

\be
\alpha > \alpha_\star \equiv \frac{\pi}{3}.
\ee
This is a surprising result. In a world described by quenched QED, massless electrons in the field of a heavy nucleus with charge $Z > 144$ would gain a mass. 

Before getting carried away it is worth observing that

(i) this result was obtained in Landau gauge, one should confirm that it remains accurate in other gauges. A clue is obtained from Eq. \ref{qed-B-2}: the factor $3$ is replaced by $3+\xi$ in general and naively (naive because we ignore $A$) $\alpha_\star$ changes to $\pi/(3+\xi)$. Clearly this is not a satisfactory situation. The problem, of course, is lack of gauge invariance induced by our simple rainbow-ladder approximation. This can be addressed by using a more complicated vertex Ansatz\footnote{The Ansatz is called the {\it Ball-Chiu vertex}. This helps but does not solve the problem. We note that Landau gauge can be regarded as preferred since $A=1$ brings the Ball-Chiu vertex near to the bare vertex.} or by solving the Schwinger-Dyson equation for the electron-photon vertex.

(ii) the mass is not given by $B$ or $B/A$ since these are Euclidean quantities. One must first rotate back to Minkowski space and then solve the equation $A(p^2)p^2 - B(p^2) = 0$ for $p^2=m^2$.

\subsection{Renormalisation}
\label{renormalisation}

Except for a few warnings, we have avoided the issue of renormalisation thus far.  This is an enormous topic and we cannot possibly do it justice here. Nevertheless, it is important to see how renormalisation fits into the functional methodology and Schwinger-Dyson equations. Thus a quick review is in order.

Renormalisation consists of three steps: 

A. {\it regulate}

B. {\it renormalise}

C. {\it remove scheme dependence}.

\subsubsection{Regulation Schemes}
Field theories are rendered finite by introducing a parameter. This can be done in any way, preferably in a way that preserves the symmetries of the action\footnote{This is not mandatory: as long as a symmetry breaking regulator does not change the universality class of the theory, the continuum limit can be recovered.}
Some popular regulation schemes are

(i) {\it dimensional regularisation}. Momentum integrals are made finite by computing in a complex dimension $d$. Continuing to $d=4$ then reveals infinities of the type $1/(d-4)$. This is the most popular regulation scheme due to its (relative) simplicity and symmetry preserving nature.

(ii) {\it Pauli-Villars}. Integrals are made finite by introducing fictitious particles that contribute with opposite sign. This scheme also preserves symmetries but can become cumbersome. It is useful in dealing with ambiguities in the definition of $\gamma_5$.

(iii) {\it lattice}. A spacetime discretisation is employed. If the lattice spacing is $a$ and the lattice size is $La$ this cuts off ultraviolet momentum integrals at $1/a$ and infrared integrals at $1/(La)$. It is very useful for numerical methods.

(iv) {\it momentum cutoff}. Simply cut off momentum integrals. This scheme violates translation invariance symmetry, which can sometimes be inconvenient (but is not a show stopper).

\subsubsection{Renormalisation Schemes}

Recall that the bare parameters of the theory are not physical. Their values must be adjusted so that the predictions of the theory agree with experiment. This is true for a completely finite theory and remains true for field theories with divergences, with the simple extension that bare parameters can be infinite when the regulator is removed. Thus, infinities are absorbed into the bare parameters of the theory. One would expect that this is generally impossible since infinities of many types can be generated by loop diagrams, while there are only a few parameters present in any theory. Remarkably, the program is possible in a wide class of field theories, which are hence called {\it renormalisable} field theories.  Nonrenormalisable theories require the introduction of an ever growing list of  parameters with which to absorb infinities\footnote{In the old days this was regarded as anathema. These days it is acceptable to deal with nonrenormalisable field theories if the parameter list can be ordered in some way, for example by energy. Theories in this class are called {\it effective field theories.}}.

In practice `absorb infinities' means define a set of {\it renormalisation conditions} that the theory must obey. For example, the mass of the electron should be 0.511 MeV and the value of $s d \sigma/d \Omega (e^+e^- \to \gamma\gamma)$ should be 25 GeV$^2$ nb/sr at $\cos \theta_{CM} = 0.28$ and $E_{CM} = 29$ GeV. Renormalisation conditions are chosen to match the problem at hand, and can sometimes be quite abstract. 

In general there must be as many renormalisation conditions as there are monomials in the theory. For $\varphi^4$ theory this means three conditions. These specify the couplings $m_0$, $\lambda_0$, and the strength of the field (corresponding to the term $1/2 (\partial \varphi)^2$). Lastly, all renormalisation schemes introduce another scale into the theory, called the {\it renormalisation scale} or the {\it subtraction point}.

(i) {\it momentum subtraction (MOM)}. Fix the residue of the propagator to be unity. Fix the pole of the propagator to the physical mass, $\Gamma^{(2)}(p^2=m^2) =0$. Fix the scattering amplitude at one point, eg, ${\cal A} = -i\lambda$ at $s=t=u=4/3 m^2$. In order, these conditions fix $\varphi_0$, $m_0$, and $\lambda_0$. MOM is the most transparent scheme for scattering processes involving physical particles\footnote{A physical particle is one that can reach a detector, like an electron. A quark is not a physical particle.}.

(ii) {\it minimal subtraction (MS)}. Directly absorb infinities in bare parameters.

(iii) {\it modified minimal subtraction ($\overline{\rm MS}$)}. Like MS but absorb some constants as well.

(iv) {\it fastest apparent convergence (FAC)}. Choose the renormalisation scale and scheme so that the last term in the perturbative expansion of a physical quantity is zero.

(v) {\it minimal sensitivity (PMS)}.  Choose the renormalisation scale and scheme so that the prediction for a physical quantity is maximally insensitive to variations in the scale.

The appearance of the renormalisation scale can be obscure. For example, in MOM one can set the scattering to a measured value at a general energy, $M$. In dimensional regularisation the coupling must be replaced with $\lambda_M = \lambda_0 M^{(4-d)}$ to obtain the correct dimensions (see Eq. \ref{unitsEq}). Once the renormalisation conditions have been imposed it is possible to remove the regulator scale $\Lambda$, leaving only the renormalisation scale $M$.

\subsubsection{The Renormalisation Group}

It is rather dispiriting to work hard and end up with an expression for a scattering cross section that depends on an arbitrary scale $M$.  Naively, your result can be any numerical value and nothing has been learned. But things are not so bad. Recall that $m$ and $\lambda$ have been determined by comparing to experiment at the scale $M$. Thus we imagine $m = m(M)$ and $\lambda= \lambda(M)$. Predictability arises because changes in $M$ lead to commensurate changes in $m$ and $\lambda$ such that the predictions remain invariant. This behaviour is called the renormalisation group and can be determined by imposing

\be
\frac{d}{d M} ({\rm physical \ quantity)} = 0,
\ee
called the {\it renormalisation group equation}. We have seen the application of this method in section \ref{evaluating}.

If all of this is getting overwhelming, there is another way to think of renormalisation. Regulate a theory with a scale, $\Lambda$. Fix $\Lambda$ and evaluate several physical quantities and adjust the parameters of the theory so that the predictions agree with experiment. Call these
$m_0(\Lambda)$ and $\lambda_0(\Lambda)$. Change $\Lambda$ to $\Lambda'$ and repeat. You will trace out the functions $m_0(\Lambda)$ and $\lambda_0(\Lambda)$ which reproduce experiment. Now compute some other quantities using these functions. If the theory is renormalisable, they will be stable as $\Lambda$ is varied. And if your theory is right and renormalisable it will not matter which observables you use to determine $m_0$ and $\lambda_0$.

\subsubsection{Renormalising the Effective Potential}
\label{renormV}

A scheme like MOM fits reasonably well into the functional formalism since it imposes conditions on the two- and four-point functions. To make contact with the effective potential we perform a derivative expansion (Eq. \ref{derExpansionEq})

\be
\Gamma(\varphi) = \int d^dx\, \left( - V_{eff}(\varphi) + \frac{1}{2} Z_{eff}(\varphi) \partial_\mu \varphi \partial^\mu \varphi  + \ldots\right).
\ee
Thus

\be 
V_{eff}(\varphi) = V_0 + V_2 \varphi^2 + V_4 \varphi^4 + \ldots
\ee
with a similar equation for $Z_{eff}$. With this expansion the 
three MOM conditions are analogous to\footnote{The exact mapping would require Fourier transforming $V_2$, $V_4$, and $Z_0$. This prescription is a little simpler.} 

\begin{eqnarray}
Z_{eff}(0) &=& Z_0 = 1 \nonumber \\
\frac{d^2}{d\varphi^2} V_{eff}|_{\varphi=0} &=& 2\, V_2 = \tilde m^2 \nonumber \\
\frac{d^4}{d\varphi^4} V_{eff}|_{\varphi=0} &=& 4!\, V_4 = \tilde \lambda.
\end{eqnarray}
The tildes indicate that these renormalisation conditions need not refer to $m$ and $\lambda$ of MOM.
Alternatively, if the limit $m\to 0$ is desired it is useful to use

\be
\frac{d^4}{d\varphi^4} V_{eff}|_{\varphi=M} = \lambda_M
\ee
since this avoids infrared divergences that appear in the effective potential in the massless limit. This condition was used to derive the renormalised effective potential given in Eq. \ref{VeffRenormEq}.

\subsubsection{Renormalising Schwinger-Dyson Equations}

The first task in renormalising  Schwinger-Dyson equations is regulating. 
Since one must employ numerical methods dimensional regulation is out of the question. This leaves Pauli-Villars, which is awkward, and momentum cutoff. Implementing a cutoff is numerically simple, but one must remember that it violates translation invariance and therefore care must be taken. For example, the computation of the vacuum polarisation tensor (the photon self energy) with a momentum cut off, $\Lambda$,  generates a term proportional to $\Lambda^2 g^{\mu\nu}$. This term violates gauge invariance, but can easily be isolated and does not cause fundamental problems.

There are several ways to impose renormalisation  conditions on the Schwinger-Dyson equations. If you are familiar with it, using the perturbative method of counterterms works well. Another method involves forming judicious subtractions. Consider, for example, the two point function of $\varphi^4$ theory given in Eq. \ref{F2-Eq}. We require that the pole location is at the physical mass,
$p^2 = m^2$, which implies

\be
F(p^2=m^2) = 0.
\ee
Now subtract zero from both sides of \ref{F2-Eq}:

\begin{eqnarray}
F(p^2) &=& p^2 - m_0^2 -i \frac{\lambda_0}{2} \int \frac{d^dq}{(2\pi)^d}\, \frac{1}{F(q^2)} + \ldots - \left[ m^2-m_0^2  -i \frac{\lambda_0}{2} \int \frac{d^dq}{(2\pi)^d}\, \frac{1}{F(q^2)} + \ldots\right] \nonumber \\
&=& p^2 - m^2 +i \frac{\lambda_0}{6} \int \frac{d^d\ell_1}{(2\pi)^d} \frac{d^d\ell_2}{(2\pi)^d} \, \frac{1}{F(\ell_1^2)}\frac{1}{F(\ell_2^2)}\cdot \nonumber \\
&& \cdot \left[\frac{\Gamma(\ell_1,\ell_2,p-\ell_1-\ell_2,-p)}{F((\ell_1+\ell_2-p)^2)} -
\frac{\Gamma(\ell_1,\ell_2,p-\ell_1-\ell_2,-p)}{F((\ell_1+\ell_2-p)^2) }\vert_{p^2=m^2}\right].
\label{FrenormEq}
\end{eqnarray}
Notice that the `bubble' diagram drops out of the equation. This makes sense, it only contributed a constant to $F$, which can immediately be absorbed into the definition of $m$. If we had truncated the Schwinger-Dyson equation at this point (ie, not included the second diagram) we would be done. 

Take $d=4$ and consider the degree of divergence of Eq. \ref{FrenormEq}. The four point function is dimensionless and so can at worst diverge like $\log\Lambda$, so we ignore it. At worst $F(\ell^2) \sim \ell^2$ so in the ultraviolet region the integral behaves like

\be
\int d^4\ell_1 d^4 \ell_2\, \frac{1}{\ell_1^2}\frac{1}{\ell_2^2} \cdot \left[\frac{p^2-m^2}{(\ell_1+\ell_2)^4}\right] \sim (p^2-m^2) \, \log \Lambda.
\ee
The degree of divergence has been knocked down from $\Lambda^2$ to $\log \Lambda$. Our expression for $F$ now looks like

\be
F(p^2) = (p^2-m^2) G(\Lambda/p) 
\label{GEq}
\ee
with 
\be
G(\Lambda/p) \sim \lambda^2 (\log \frac{\Lambda^2}{p^2} + {\rm finite}).
\ee
The logarithmic divergence is eliminated by imposing the residue condition

\be
\frac{d}{dp^2} F(p^2)\vert_{p^2=m^2} = 1.
\ee
Thus $G(\Lambda/m)$ must be unity. Subtract $0 = G(\Lambda/m)-1$ again to obtain the renormalised Schwinger-Dyson equation for the two point function:

\be
F(p^2) = (p^2-m^2)\cdot \left[G(\Lambda/p) - G(\Lambda/m) + 1\right]
\ee
where $G$ is determined by Eqs. \ref{FrenormEq} and \ref{GEq}. Notice that the dependence on the cutoff is eliminated since the $\Lambda$-dependent portion of the term in brackets is

\be
\log \frac{\Lambda^2}{p^2} - \log \frac{\Lambda^2}{m^2} = \log \frac{m^2}{p^2}.
\ee

A single subtraction will suffice to render the equation for $\Gamma^{(4)}$ finite and will replace references to $\lambda_0$ with $\lambda$. Since $\varphi^4$ theory is renormalisable in four dimensions, this is all that is required to make all other Schwinger-Dyson equations finite.

\subsection{Truncation Schemes}
\label{truncation}

Although Schwinger-Dyson equations are equivalent to the effective potential, and hence to the generating functional, and hence to the quantum field theory, and they permit resumming infinitely many Feynman diagrams, they still form an infinite hierarchy of equations. Thus some form of truncation is required. Truncation amounts to searching for a small parameter. Small parameters that have been employed in the literature are

(i) {\it coupling constant.}   If $\lambda$ is small diagrams with many vertices are suppressed and the Schwinger-Dyson equations can be ordered. Of course this is equivalent to standard perturbation theory, and one is advised to use well-established perturbative methods in this case.

(ii) {\it $1/N$, $1/N_c$.}   If the number of fields (either labelled by an external index or an internal degree of freedom such as flavour or colour) is large, diagrams that contain fewer fields are suppressed, supplying an ordering mechanism.

(iii) {\it classical limit.}   If classical physics dominates the system one can order diagrams in powers of $\hbar$. This method generates the {\it loop expansion} of the effective action.

(iv) {\it high density or temperature.}  If a system is coupled to a heat bath, thermodynamic quantities such as temperature and density (high or low) will sometimes provide a useful ordering principle. This method is used, for example, in the hole-line expansion for low density nuclear matter.

(v) {\it infrared dominance.}  Many systems are dominated by long range interactions. Models of these systems typically contain (naive) infrared divergences. In these cases, ordering diagrams according to their degree of infrared divergence can yield a useful truncation scheme. This  method was employed in the resolution of the infrared divergence problem in Fermi gasses by Brueckner and Gell-Mann.

(vi) {\it guess.}
Guesses typically take the form of truncating the Schwinger-Dyson equations at low $n$ where $n$ refers to the $n$-point function. While this is an organisational scheme, it is not a truncation scheme.  Unfortunately this method is the one most commonly employed in the literature. It is not necessarily bad, but correctly implementing it requires that one check the efficacy of the guess and then confirm that the guess is robust.

\subsection{Numerical Methods}
\label{numerical}

In general it is impossible to solve Schwinger-Dyson equations analytically and one must resort to numerical methods. Unfortunately there is no general method for solving nonlinear coupled integral equations and one must live by one's wits. Strategies that have been attempted are

\vskip .3 cm
\noindent
A) reduce the functional problem to an algebraic problem. This is done in one of two ways

(i) expand functions in terms of a convenient basis: $A(p) = \sum_{i} c_i T_i(p)$, where hopefully the properties of $T$ allow some integrals to be performed exactly (this is only the case for very simple kernels, which means highly truncated Schwinger-Dyson equations). A benefit of this approach is that it is simple to evaluate quantities like $A(|p+q|)$, which are often required.  A demerit is that the basis may be too restrictive to permit an accurate solution (for example, obtaining the oscillatory solutions of rainbow-ladder QED would be very difficult with  polynomial basis functions).

(ii) place the functions on a momentum grid: $A(p) \to A_i = A(p_i)$. This method has the benefit that integrals are easy to obtain (numerically) and that the solution is not confounded by basis assumptions. Alternatively, it makes evaluation of $A$ on non-grid points difficult.

\vskip .3 cm
\noindent
B) solve the resulting algebraic equations for the $A_i$ or $c_i$. This is a difficult problem in general. Possible approaches are

(i) {\it iteration} one could attempt this by simply iterating the gap equations, using the LHS from the previous iteration to evaluate the RHS. This is the simplest possible procedure but suffers from instability, as is easily verified in one dimension.

(ii) {\it iterative Newton-Raphson}  stability can be improved by linearising the gap equations to obtain a new estimate of the root. Thus if one wishes to solve 

\begin{equation}
x_i = f_i(\{x\})
\label{gap-discrete}
\end{equation}
expand about an assumed solution and neglect higher order terms
\be
x_i + \delta x_i = f_i(\{x\}) + \frac{\partial f_i}{\partial x_j}|_{x_i} \delta x_j
\ee
and solve for the corrections
\be
\sum_j\left( - \delta_{ij} + \frac{\partial f_i}{\partial x_j}|_{x_i} \right) \delta x_j = x_i - f_i(\{x\}).
\ee
One then sets $x_i = x_i + \delta x_i$ and iterates. In general this algorithm converges to a sensible solution more often than simple iteration.

(iii) {\it minimisation} construct the function $G(\{ x \}) = \sum_i (x_i - f_i(\{ x \}))^2$ and minimise it. This can be a powerful method but suffers from the bane of all minimisation problems, that of finding the global minimum in a multidimensional sea of local minima. In fact, $G$ often has local minima in the neighbourhood of zeroes of the functions $f_i$. Unfortunately, these need not be at all close to solutions to Eq. \ref{gap-discrete}.

In fact all of the techniques discussed here will fail miserably unless one starts {\it very close} to the solution. Thus it is vital that the practitioner not abandon theoretical investigations too early. One must carefully track and deal with singularities in the equations, understand asymptotic behaviour, and develop decent analytic approximations to have any hope of obtaining reliable numerical solutions.
And it is best to be prepared for a lot of heartbreak.

\section{Conclusions}

\subsection{Summary}

I hope that this short introduction has convinced you that functional methods are a very powerful and flexible tool for examining quantum field theory. At the least, they are more compact than the canonical approach, although somewhat more `formal'. Dealing with advances topics such as curved spacetime, anomalies, or Faddeev-Popov gauge fixing is also much easier in the functional approach. Lastly, we have seen how the familiar machinery of quantum mechanics can be carried over to quantum field theory in the Schr\"odinger functional approach. 

Schwinger-Dyson equations are a powerful method for organising and summing infinitely many Feynman diagrams. As such they represent one of the few ways to obtain nonperturbative information on a field theory. And we have seen that in some cases it is very easy to obtain such information. The equivalence to other nonperturbative methods, the Bogoliubov canonical transformation and the Gaussian variational Ansatz, has also been shown. However, it is clear that the Schwinger-Dyson method is much more general than either of these. It is also easy to generalise the formalism to finite temperature and density.

Nevertheless, problems lurk. The technical challenge in solving coupled nonlinear integral equations over many variables can be daunting. More fundamentally, deciding how to truncate the equations is vital and, in most cases, not obvious. There is room for improvement here!

\subsection{Where to go from here}

These notes are an introduction to an enormous field with applications across physics; much has been omitted!
For starters fermions were almost entirely neglected due to  spacetime limitations. A large literature exists on properties of path integrals, since their definitions are subtle. Applications to the Standard Model and attempts to go beyond the Standard Model abound. Thus, I include a small bibliography to help explore these topics.

\subsubsection{Original Articles}

P.A.M. Dirac, {\sl Physikalische Zeitschrift der Sowjetunion}, {\bf 3}, 64 (1933). The germ of the path integral.

R.P. Feynman, {\sl Rev. Mod. Phys.}, {\bf 20}, 367 (1948). Feynman develops the path integral.

R.P. Feynman and A.R. Hibbs, {\sl Quantum Mechanics and Path Integrals} (McGraw-Hill, New York, 1965). Apparently this book has become a collector's item, as new copies are going for \$600 on amazon.com! Look for the rare ones that forgot the word `Path' in the title.

F.A. Berezin, {\sl The Method of Second Quantisation}, (Academic Press, New York, 1966).
Introduced the use of Grassmann variables to deal with fermions.

S. Coleman and E. Weinberg, {\sl Phys. Rev. D} {\bf 7}, 1888 (1973). Effective potentials and hidden symmetry.

\subsubsection{Books on Functional Methods}

R.J. Rivers, {\sl Path Integral Methods in Quantum Field Theory},  (Cambridge University Press, New York ,1990).

L.S. Schulman, {\sl Techniques and Applications of Path Integration}, (Dover, London, 2005).

M.S. Swanson, {\sl Path Integrals and Quantum Processes}, (Academic Press, New York, 1992). Nope. Not related.

\subsubsection{Other Books}

D.J. Amit, {\sl Field Theory, the Renormalization Group,  and Critical Phenomena} (World Scientific, Singapore, 1989). A nice introduction, with a rapid approach to advanced topics, applications tend to condensed matter physics.

P. Ramond, {\sl Field Theory: a Modern Primer}, (Benjamin/Cummings Publishing, Reading, MA, 1981). A very nice pedagogical introduction to field theory and the functional method.

V.A. Miransky, {\sl Dynamical Symmetry Breaking in Quantum Field Theories}, (World Scientific, Singapore, 1993).
Discusses dynamical symmetry breaking with a focus on chiral symmetry breaking in a number of systems, including superconductivity, the Nambu-Jona-Lasinio model, QCD, the Standard Model, and technicolour. Chapter 7 is an excellent introduction to the effective action.

C. Itzykson and J.-B. Zuber, {\sl Quantum Field Theory} (McGraw-Hill, New York, 1980). The standard reference. Unfortunately, functional methods are scattered throughout the text. Look in sections
5-1, 6-2-2, 9-1, 10-1, and 11-2-2. Treatment of the Schwinger-Dyson equations is light.

V.N. Popov, {\sl Functional Integrals and Collective Excitations} (Cambridge University Press, New York, 1987). Applications to condensed matter physics; does not make heavy use of the functional approach.

M.E. Peskin and D.V. Schroeder, {\sl An Introduction to Quantum Field Theory} (Westview Press, 1995). See chapter 11.

\subsubsection{Review Articles}

Applications to QCD and other field theories are reviewed in many articles. The following  will get you nearly up to date.

  R.~Alkofer and L.~von Smekal,
  {\sl The infrared behavior of QCD Green's functions: Confinement, dynamical
  symmetry breaking, and hadrons as relativistic bound states}, 
  Phys.\ Rept.\  {\bf 353}, 281 (2001).

 C.~S.~Fischer,
 {\sl Infrared properties of QCD from Dyson-Schwinger equations},
 J.\ Phys.\ G {\bf 32} (2006) R253.

  C.K. Kim and S. K. You, {\sl The Functional Schr\"odinger Picture Approach to Many-Particle Systems}, cond-mat/0212557.

 P.~Maris and C.~D.~Roberts,
 {\sl Dyson-Schwinger equations: A tool for hadron physics},
 Int.\ J.\ Mod.\ Phys.\  E {\bf 12}, 297 (2003).

 M. Pennington, {\sl Swimming with Quarks}, hep-ph/0504262. 

  C.~D.~Roberts and S.~M.~Schmidt,
  {\sl Dyson-Schwinger equations: Density, temperature and continuum strong QCD},
  Prog.\ Part.\ Nucl.\ Phys.\  {\bf 45}, S1 (2000).

  C.~D.~Roberts and A.~G.~Williams,
  {\sl Dyson-Schwinger equations and their application to hadronic physics},
  Prog.\ Part.\ Nucl.\ Phys.\  {\bf 33}, 477 (1994).

\section*{Acknowledgments}
These notes are based on lectures given at the XI Hadron Physics Summer School in Maresias, Brazil and at the 25th Hampton University Graduate Studies Program at Jefferson Lab. The author is grateful to Gast\~{a}o Krein, Marina Nielson, Fernando Navarra, and Tereza Mendes for warm hospitality extended to him during his stay in Brazil and to Rolf Ent and Vadim Guzey for the invitation and assistance at HUGS.

\end{document}